\documentclass[11pt,a4paper]{article}
\usepackage{amsthm,amssymb,latexsym,epsfig}
\newcommand{\be}[1]{\begin{equation}\label{#1}}
\newcommand{\ba}[1]{\begin{eqnarray}\label{#1}}
\newcommand{\ee}{\end{equation}}
\newcommand{\ea}{\end{eqnarray}}
\newcommand{\non}{\nonumber\\\rule{0pt}{30pt}}
\newcommand{\nona}[1]{\nonumber\\\rule{0pt}{#1pt}}
\newcommand{\num}{\\\rule{0pt}{20pt}}

\newcommand{\dis}{\displaystyle}
\newcommand{\eq}[1]{(\ref{#1})}
\newcommand{\tr}{\mathop{\rm tr}}
\newcommand{\Res}{\mathop{\rm Res}}
\newcommand{\bra}[1]{\langle{#1}|}
\newcommand{\ket}[1]{|{#1}\rangle}
\newtheorem{thm}{Theorem}[section]
\newtheorem{prop}{Proposition}[section]
\newtheorem{lemma}{Lemma}[section]
\newtheorem{cor}{Corollary}[section]
{\theoremstyle{remark}
\newtheorem{rem}{Remark}[section]}
\def\qed{\hfill\nobreak\hbox{$\square$}\par\bigbreak}
\makeatletter
\@addtoreset{equation}{section}
\makeatother

\begin{document}
\begin{flushright}
LPENSL-TH-01/02\\
\end{flushright}
\par \vskip .1in \noindent

\vspace{24pt}

\begin{center}
\begin{LARGE}
{\bf Spin-spin correlation functions of the $XXZ$-$\scriptstyle{\frac{1}{2}}$
Heisenberg chain in a magnetic field}
\end{LARGE}

\vspace{50pt}

\begin{large}

{\bf N.~Kitanine}\footnote[1]{Graduate School of Mathematical
Sciences, University of Tokyo, Japan,
kitanine@ms.u-tokyo.ac.jp\par
\hspace{2mm} On leave of absence from Steklov Institute at
St. Petersburg, Russia},~~
{\bf J.~M.~Maillet}\footnote[2]{ Laboratoire de Physique, UMR 5672 du CNRS,
ENS Lyon,  France,
 maillet@ens-lyon.fr},~~
{\bf N.~A.~Slavnov}\footnote[3]{ Steklov Mathematical Institute,
Moscow, Russia, nslavnov@mi.ras.ru},~~
{\bf V.~Terras}\footnote[4]{NHETC, Department of Physics and Astronomy,
Rutgers University, NJ, USA, vterras@physics.rutgers.edu \par
\hspace{2mm} On leave of absence from LPMT, UMR 5825 du CNRS,
Montpellier, France}
\end{large}

\vspace{80pt}

\centerline{\bf Abstract} \vspace{1cm}
\parbox{12cm}{\small Using algebraic Bethe ansatz and the solution of the
quantum inverse scattering problem, we compute compact
representations of the spin-spin correlation functions of the
$XXZ$-$1 \over 2$ Heisenberg chain in a magnetic field. At lattice
distance $m$, they are typically  given as the sum of $m$ terms.
Each term $n$ of this sum, $n = 1, \dots m$, is represented in the
thermodynamic limit as a multiple integral of order $2n+1$; the
integrand  depends on the distance as the power $m$ of some simple
function. The root of these results is the derivation of a compact
formula for the multiple action on a general quantum state of the
chain of transfer matrix operators for arbitrary values of their
spectral parameters.}
\end{center}

\newpage

\section{Introduction}

\bigskip

The main challenging problem in the field of quantum integrable
models is to compute exact and manageable expressions for their
correlation functions. This issue is of great importance, not only
from a  theoretical view point  but also for applications to
relevant physical situations.

\bigskip

The archetype of quantum integrable lattice models is provided by
the $XXZ$ spin-$1 \over 2$ Heisenberg chain in a magnetic field,
\be{IHamXXZ} H=\sum_{m=1}^{M}\left(
\sigma^x_{m}\sigma^x_{m+1}+\sigma^y_{m}\sigma^y_{m+1}
+\Delta(\sigma^z_{m}\sigma^z_{m+1}-1)\right)-hS_z, \ee
where
\be{ISz} S_z=\frac{1}{2}\sum_{m=1}^{M}\sigma^z_{m},\qquad
[H,S_z]=0. \ee
Here $\Delta$ is the anisotropy parameter, $h$ an external
classical magnetic field, and  $\sigma^{x,y,z}_{m}$ denote the
usual Pauli matrices acting on the quantum space at site $m$ of
the chain. The simultaneous reversal of all spins is equivalent to
the change of the sign of the magnetic field, therefore it is
enough to consider the case $h\ge 0$. In the thermodynamic limit
$M\to\infty$ and at zero magnetic field, the model exhibits three
different regimes depending on the value of $\Delta$: for $\Delta
\le -1$, the model is ferromagnetic; for $-1 < \Delta \le 1$, the
model has a non degenerated anti-ferromagnetic ground state, and
the spectrum is gapless (massless regime); for $\Delta > 1$,
the ground state is twice degenerated with a gap in the spectrum
(massive regime).

\bigskip

Although the method to compute eigenstates and energy levels goes
back to H. Bethe in 1931 \cite{Bet31,Orb58,Wal59,YanY66a}, the
knowledge of its spin correlation functions has been for a long
time restricted to the free fermion point $\Delta = 0$, a case for
which nevertheless tremendous works have been necessary to obtain
full answers
\cite{LieSM61,Mcc68,WuMTB76,McCTW77,SatMJ78,ColIKT93}.

However, at zero temperature and for zero magnetic field, {\em
multiple integral representations of  elementary blocks of the
correlation functions} (see definition below) have been obtained
from the $q$-vertex operator approach (inspired from the corner
transfer matrix technique) in the massive regime $\Delta > 1$ in
1992 \cite{JimMMN92}, and conjectured in 1996 \cite{JimM96} for the
massless regime $-1 < \Delta \le 1$ (see also \cite{JimML95}). A
proof of these results together with their extension to non-zero
magnetic field has been obtained in 1999 \cite{KitMT99,KitMT00}
for both regimes using algebraic Bethe ansatz
\cite{FadST79,TakF79} and the actual resolution of the so-called
quantum inverse scattering problem \cite{KitMT99,MaiT00}.


These elementary blocks for correlation functions are defined in
the following way:
\be{genabcd00} F_m(\{\epsilon_j,\epsilon'_j\})=\frac
{\bra{\psi_g}\prod\limits_{j=1}^m
E^{\epsilon'_j,\epsilon_j}_j\ket{\psi_g}}{\bra{\psi_g}\psi_g\rangle}
. \ee
Here $\ket{\psi_g}$ denotes the ground state in the massless regime or
any of the two ground states constructed by algebraic Bethe ansatz
in the massive regime; $E^{\epsilon'_m,\epsilon_m}_{m}$ are the
elementary operators acting on the quantum space ${\cal H}_m$ at
site $m$ as the $2 \times 2$ matrices
$E^{\epsilon',\epsilon}_{lk}=\delta_{l,\epsilon'}\delta_{k,\epsilon}$.
Any $n$-point correlation function can be reconstructed as a sum
of such elementary blocks.

\bigskip

To compute these elementary blocks,
the following successive problems have to be addressed
\cite{KitMT99,KitMT00}: $(i)$ determination of the ground state
$\bra{\psi_g}$, $(ii)$ evaluation of the action of the product of
the local operators $E^{\epsilon'_j,\epsilon_j}_{j}$ on this
ground state, and $(iii)$ computation of the scalar product of the
resulting state with $\ket{\psi_g}$.

For the $XXZ$ spin-$1 \over 2$ Heisenberg chain in a magnetic
field, these problems have been solved in \cite{KitMT99,KitMT00}
in the framework of algebraic Bethe ansatz. The central object of
this method is the so called quantum monodromy matrix depending on
a complex variable $\lambda$ (the spectral parameter). For the
$XXZ$ spin-$1 \over 2$ chain, it is a $2 \times 2$ matrix with
operator valued entries acting in the quantum space of states
$\cal H$:
\be{IABAT} T(\lambda)=\left(
\begin{array}{cc}
A(\lambda)&B(\lambda)\\
C(\lambda)&D(\lambda)
\end{array}\right).
\ee
The quadratic commutation relations between these four operators
are given by the Yang-Baxter algebra. It is governed by a
trigonometric $R$ matrix solving the Yang-Baxter equation. The
Hamiltonian of the chain is then contained in the commutative
family of operators generated by the transfer matrix ${\cal
T}(\lambda) = (A + D)(\lambda)$ for arbitrary values of $\lambda$.
The algebraic Bethe ansatz leads to the simultaneous
diagonalization of these transfer matrices and of the Hamiltonian.
The ground state $\bra{\psi_g}$ (resp. $\ket{\psi_g}$), as the
other eigenstates, is given as the successive action of operators
$C(\lambda_k)$ (resp. $B(\lambda_k)$) on the ferromagnetic
reference state $\bra{0}$ (resp. $\ket{0}$) with all spins up.
Namely, we have $\bra{\psi_g} = \bra{0} \prod_k C(\lambda_k) $ and
$\ket{\psi_g} = \prod_k B(\lambda_k) \ket{0}$ for a particular set
of spectral parameters $\{\lambda_k\}$ solving the Bethe
equations.

To evaluate the action of local operators on this state, the
strategy is to  imbed them in the Yang-Baxter algebra of $T$
matrices by solving the quantum inverse scattering problem (see
\cite{KitMT99,MaiT00} for details) as
\be{ET}
    E^{\epsilon'_j,\epsilon_j}_j=\prod_{k=1}^{j-1}
       \bigl( A+D \bigr) ({\textstyle\frac\eta2})\
       T_{\epsilon_j,\epsilon'_j}({\textstyle\frac\eta2})\
       \prod_{k=1}^{j} \bigl( A+D \bigr)^{-1} ({\textstyle\frac\eta2}),
\ee
where $\cosh \eta = \Delta$. Then, using the Yang-Baxter algebra,
one can reduce any elementary blocks of the correlation functions
to multiple sums of scalar products of some states with
$\ket{\psi_g}$. Each of these scalar products can be computed as
the ratio of two explicit determinants \cite{Sla89,KitMT99}. In
the thermodynamic limit, these multiple sums lead to $m$ fold
integrals over contours $C_j^h$ which depend on the value of $j$,
on the regime considered and also on the value of the magnetic
field. The answer can be written generically as \cite{KitMT00}
\be{Fmh}
  F_m (\{\epsilon_j,\epsilon'_j\})=
    \prod\limits_{j=1}^{m} \int _{C_j^h} d\lambda_{j}\ \
    \Omega_m(\{\lambda\},\, \{\epsilon_j,\epsilon'_j\})\ S_h(\{\lambda\}).
\ee
Here $\Omega_m (\{\lambda\},\, \{\epsilon_j,\epsilon'_j\})$ is a
purely algebraic quantity, which in particular does not depend on
the regime nor on the magnetic field. In contrast,
$S_h(\{\lambda\})$ is a functional of the  density function
$\rho_h(\lambda)$ solution of the Lieb equation describing the
ground state \cite{YanY66a}, and hence depends both on the regime
and on the value of the magnetic field $h$.

It is remarkable that, for zero magnetic field, the two  multiple
integral representations resulting from the $q$-vertex operator
approach and from Bethe ansatz are identical: contours and
integrands coincide. It would be very desirable to understand this
intriguing fact directly at the operator level. Note that for non
zero magnetic field, the quantum affine symmetry used in the
$q$-vertex operator approach is broken, and no result is known up
to now from this method in this case.

\bigskip

In principle, any $n$-point correlation function can be obtained
from these elementary building blocks. One should note however
that, although these formulas are quite explicit, the actual
analytic computation of these multiple integrals is missing up to
now. Moreover, the evaluation of correlation functions of physical
relevance, as for example the spin-spin correlation functions at
distance $m$ on the lattice like
$\langle\sigma_1^+\sigma_{m+1}^-\rangle$, is a priori quite
involved. Indeed, the identity
\be{s+s-} \bra{\psi_g} \sigma^+_1\,
\sigma^-_{m+1}\ket{\psi_g}\equiv\bra{\psi_g}E^{12}_1\prod\limits_{j=2}^{m}
(E^{11}_j + E^{22}_j)\, E^{21}_{m+1}\ket{\psi_g} \ee
shows that the corresponding spin-spin correlation function is
actually given as a {\em sum of $2^{m-1}$ elementary blocks}. So,
the number of terms to sum up grows exponentially with $m$, making
the problem of asymptotic behavior at large distance extremely
difficult to solve in these settings from the present knowledge of
the elementary blocks \eq{Fmh}. In the language of algebraic Bethe
ansatz, and using the solution of the quantum inverse scattering
problem,  this question amounts to the computation of the
following average value:
\be{CA+DB} \bra{\psi_g}C({\textstyle\frac\eta2})\, (A + D)^{m-1}
({\textstyle\frac\eta2})\, B ({\textstyle\frac\eta2}) \ket{\psi_g} .
\ee
Hence, to obtain manageable (re-summed) formulas for spin-spin
correlation functions, and to avoid the computation of the above
sum of $2^{m-1}$ terms,  we need to derive a compact expression
for the action of the shift operator $(A + D)^{m-1}({\eta \over
2})$ (from site $1$ to site $m+1$) on arbitrary states.

\bigskip

The main purpose of this paper is to give a solution to this
problem in the framework of algebraic Bethe ansatz, and to apply it
to the evaluation of spin-spin correlation functions.

\bigskip

In fact, for later use, we will solve an even more general
question: the evaluation of a compact formula for the multiple
action of transfer matrix operators $(A + D)(x_{\alpha})$, for any
set of spectral parameters $x_{\alpha}$, on arbitrary quantum
states (a priori not eigenstates) of the $XXZ$ model (see
Proposition \ref{act-t}).  This leads to {\em the evaluation of
the spin-spin correlation functions at lattice distance $m$ as the
sum of only $m$ terms (instead of $2^{m-1}$)}, the $n^{th}$ term
in the sum being expressed in the thermodynamic limit as a
multiple integral of order $2n+1$, for $n = 1, \dots, m$ (see
Proposition \ref{sz-sz}, Proposition \ref{sp-sm}). For the two
point correlation functions, a typical form of the result is
\eq{Os+s-},
\ba{InewGF} &&{\dis\hspace{-2mm}
\langle\sigma^{\alpha}_1\sigma_{m+1}^{\beta}\rangle=
\sum_{n=0}^{m-1}\ \oint\limits_{C_z}d^{n+1} z
\int\limits_{C_{\lambda}} d^{n}\lambda \int\limits_{C_{\mu}}
d^{2}\mu\ [f (\{\lambda,z\})]^m\ \Gamma^{\alpha
\beta}_n(\{\lambda,\mu,z\})\ S_h(\{\lambda,z\})}. \ea
Here, $\alpha,\beta = x,y,z$, and the functions  $\Gamma^{\alpha
\beta}_n(\{\lambda,\mu,z\})$, $f(\{\lambda,z\})$ are purely
algebraic quantities, which do not depend on the regime nor on the
magnetic field; the integration contours $C_z, C_{\lambda},
C_{\mu}$ and the functional $S_h$ of the density function
describing the ground state (evaluated at points $\lambda$ and
$z$) depend both on the regime and on the value of the magnetic
field.

In this formula, one can interpret the integrals over the $\lambda$
and $z$ variables to be  generically associated to the shift
operator from site $1$ to site $m+1$, while the integrals over the
variables $\mu$ correspond to the contribution of the operators
$\sigma^{\alpha}, \sigma^{\beta}$.

Hence this method provides us with an effective re-summation of
the previous $2^{m-1}$ elementary blocks, although we found it
more convenient and general to work it out at the operator level
and in the algebraic Bethe ansatz framework. In particular, all
our considerations are valid for the finite lattice case. We also
believe that it can be applied to many other models for which
multiple integral representations of elementary blocks of
correlation functions are known, like for example the integrable
higher spin Heisenberg chains \cite{Kit01}.

Moreover, it should be stressed here that, for each term $n$ of
this sum, {\em the distance appears now explicitly in the
integrand merely as the power $m$ of some function
$f(\{\lambda,z\})$} of the integration variables. This feature,
which is the result of our re-summation, is obviously of great
importance for future asymptotic analysis at large $m$. Let us
finally mention that these new representations of the spin-spin
correlation functions (valid for arbitrary values of $\Delta >
-1$) lead in a simple way to the known results at the free fermion
point $\Delta = 0$; the corresponding computations will be
presented in a separate publication.

\bigskip

This article is organized as follows. In the next section, we
recall some basics about the study of the $XXZ$
spin-$\scriptstyle{\frac{1}{2}}$ Heisenberg chain in the algebraic
Bethe ansatz framework. In section \ref{CCF}, we present the list
of formulas necessary for the computation of the correlation
functions via algebraic Bethe ansatz. In section \ref{AP}, we
derive a compact formula for the multiple action of the transfer
matrix operator on an arbitrary quantum state of the chain and for
any value of the spectral parameters. It leads in section \ref{GF}
to the evaluation of the generating functional of the $\sigma^z$
correlation functions. General spin-spin correlation function at
lattice distance $m$ are given in section \ref{ssc}. Some
 perspectives are discussed in the conclusion. Lengthy computations
and/or proofs of intermediate results are presented in a set of
three appendices.

\bigskip

We dedicate this paper to the memory of our friend and colleague
A. Izergin. When we began this work two years ago, he was about to
join us, but unfortunately these plans were suddenly stopped.


\section{The $XXZ$ spin-$\scriptstyle{\frac{1}{2}}$ Heisenberg chain
\label{Intro2}}

The Hamiltonian of the cyclic $XXZ$ chain with $M$ sites is given
by \eq{IHamXXZ}. In the framework of algebraic Bethe ansatz, it
can be obtained from the monodromy matrix $T(\lambda)$, which is
in turn completely defined by the $R$-matrix. The $R$-matrix of
the $XXZ$ chain acts in the space $\mathbb{C}^2\otimes
\mathbb{C}^2$ and is equal to
\be{ABAR}
R(\lambda)=\frac{1}{\sinh(\lambda+\eta)} \left(
\begin{array}{cccc}
\sinh(\lambda+\eta)&0&0&0\\
0&\sinh\lambda&\sinh\eta&0\\
0&\sinh\eta&\sinh\lambda&0\\
0&0&0&\sinh(\lambda+\eta)
\end{array}
\right),\qquad \cosh\eta=\Delta.
\ee
It is a trigonometric  solution of the Yang-Baxter equation.
Identifying one of the two vector spaces of the $R$-matrix with
the quantum space ${\cal H}_m$, one defines the quantum $L$-operator at
site $m$ by
\be{ABL}
L_m(\lambda)=R_{0m}(\lambda-\eta/2).
\ee
Here $R_{0m}$ acts in $\mathbb{C}^2\otimes {\cal H}_m$. The
monodromy matrix $T(\lambda)$ is then constructed as an ordered
product of the $L$-operators with respect to all the sites of the
chain:
\be{ABAT}
T(\lambda)=\left(
\begin{array}{cc}
A(\lambda)&B(\lambda)\\
C(\lambda)&D(\lambda)
\end{array}\right)=L_M(\lambda)
\dots L_2(\lambda) L_1(\lambda).
\ee
The Hamiltonian \eq{IHamXXZ} at $h=0$ can be obtained from
$T(\lambda)$ by the trace identity
\be{ABATI}
H=2\sinh\eta\left.\frac{\partial}{\partial\lambda} \log
{\cal T}(\lambda)\right|_{\lambda=\frac{\eta}{2}}+const.
\ee
Here, the transfer matrix
\be{ABAtr}
{\cal T}(\lambda)=\tr T(\lambda)=A(\lambda)+D(\lambda)
\ee
generates a continuous set of commuting conserved quantities.  For
technical reasons it is convenient to consider the inhomogeneous
$XXZ$ model, where
\be{ABALinh}
L_m(\lambda)=L_m(\lambda,\xi_m)=R_{0m}(\lambda-\xi_m), \qquad
T(\lambda)=L_M(\lambda,\xi_M)
\dots L_2(\lambda,\xi_2)  L_1(\lambda,\xi_1),
\ee
and $\xi_m$ are arbitrary complex numbers attached to each lattice
site that are called inhomogeneity parameters. In the homogeneous
limit $\xi_m=\eta/2$, we come back to the original model
\eq{IHamXXZ}. The commutation relations between the entries of the
monodromy matrix are given  by the Yang-Baxter quadratic relation,
\be{ABARTT}
R_{12}(\lambda_1-\lambda_2)T_1(\lambda_1)T_2(\lambda_2)=
T_2(\lambda_2)T_1(\lambda_1)R_{12}(\lambda_1-\lambda_2).
\ee
The equation \eq{ABARTT} holds in the space $V_1\otimes V_2\otimes
{\cal H}$ (where $V_j\sim \mathbb{C}^2$). The matrix
$T_j(\lambda)$ acts in a nontrivial way in the space $V_j\otimes
{\cal H}$, while the $R$-matrix $R_{12}$ is nontrivial in
$V_1\otimes V_2$.

The space of states is  generated  by the action of creation
operators $B(\lambda)$ and annihilation operators $C(\lambda)$  on
the reference state $|0\rangle$ with all spins up. In the following, we will
consider general states of the form
\be{ABAES} |\psi\rangle=\prod_{j=1}^{N}B(\lambda_j)|0\rangle,
\qquad N=0,1,\dots, M, \ee
which are  eigenstates of the transfer matrix ${\cal T}(\mu)$ (and
thus of the Hamiltonian in the homogeneous case) when the
parameters $\lambda_j$ satisfy the system of Bethe equations
\be{ABABE}
\prod_{m=1}^M\frac{\sinh(\lambda_j-\xi_m)}{\sinh(\lambda_j-\xi_m+\eta)}
\cdot\prod_{k=1\atop{k\ne
j}}^{N}  \frac{\sinh(\lambda_j-\lambda_k+\eta)}
{\sinh(\lambda_j-\lambda_k-\eta)}=1,
\qquad j=1,\dots, N.
\ee
The corresponding eigenvalue $\tau(\mu,\{\lambda_j\})$ of the
operator ${\cal T}(\mu)$ (containing the energy level following
\eq{ABAR}) is
\be{ABAEV}
\tau(\mu,\{\lambda_j\})=
a(\mu)\prod_{j=1}^{N}f(\lambda_j,\mu)+d(\mu)
\prod_{j=1}^{N}f(\mu,\lambda_j).
\ee
Here and further we use abbreviated notations for certain
combinations of hyperbolic functions:
\be{ABAft}
f(\lambda,\mu)=\frac{\sinh(\lambda-\mu+\eta)}{\sinh(\lambda-\mu)}, \qquad
t(\lambda,\mu)=\frac{\sinh\eta}{\sinh(\lambda-\mu)\sinh(\lambda-\mu+\eta)}.
\ee
The functions $d(\mu)$ and $a(\mu)$ are eigenvalues of the
operators $D(\mu)$ and $A(\mu)$ on the reference state:
\be{ABAev}
d(\mu)=\prod_{m=1}^Mf^{-1}(\mu,\xi_m),\qquad a(\mu)=1.
\ee
Following the paper \cite{KitMT00}, we consider below the action
of the monodromy matrix elements on the dual state which can be
constructed similarly to \eq{ABAES} via the operators $C(\lambda)$
as
\be{ABADES}
\langle\psi|=\langle 0|\prod_{j=1}^{N}C(\lambda_j),
\qquad N=0,1,\dots, M.
\ee
Here  $\langle 0|=|0\rangle^{+}$, and \eq{ABADES}  defines a dual
eigenstate if the parameters $\lambda_j$ satisfy the same system
of  Bethe equations \eq{ABABE}.

\bigskip

Our final goal is to compute the correlation functions in the
ground state in the thermodynamic limit  $M\to\infty$. The
thermodynamics of the $XXZ$ chain was studied  in
\cite{LieSM61,YanY66a,LieL63}. Here we merely recall the formulas we need
for our study.

The ground state $|\psi_g\rangle$ of the infinite chain can be
constructed as the limit of the finite chain eigenstate
$\prod_{j=1}^{N}B(\lambda_j)|0\rangle$ for $M\to\infty$,
$N\to\infty$ and $N/M$ equal to some constant whose value
depends on the magnetic field $h$. In this limit, the Bethe
equations for the set of parameters $\{\lambda_j\}$ reduce to  the
integral Lieb equation for the ground state spectral density
$\rho_{tot}(\lambda)$:
\be{GFLiebeq} -2\pi
i\rho_{tot}(\lambda) +\int_{C} K(\lambda-\mu)\rho_{tot}(\mu)\,d\mu=
t(\lambda,{\textstyle\frac\eta2}), \ee
where
\be{GFKkern}
K(\lambda)=\frac{\sinh2\eta}
{\sinh(\lambda+\eta)\sinh(\lambda-\eta)}.
\ee
The integration contour $C = [-\Lambda_h,\Lambda_h]$ in
\eq{GFLiebeq} depends on the regime considered.  In the massless
case $-1 < \Delta \le 1$, the contour $C$ is an interval of the
real axis and the parameter $\eta$ is imaginary: $\eta=-i\zeta$,
$\zeta>0$. In particular, at $h \to 0$, $\Lambda_h \to \infty$,
and the Lieb equation can be solved explicitly (see
\eq{GFsolmasles}). For $\Delta>1$ ($\eta<0$) the limits
$\pm\Lambda_h$ are imaginary, which means that the integral in
\eq{GFLiebeq} is taken over an interval of the imaginary axis. At
$h=0$, $\Lambda_h = -i\pi/2$ and the solution of the Lieb equation
is given in terms of theta-functions (see \eq{GFsolmasles}).

For technical purposes, we also introduce the inhomogeneous
density $\rho(\lambda,\xi)$ as the solution of the integral
equation
\be{GFinteqinh} -2\pi i\rho(\lambda,\xi) +\int_{C}
K(\lambda-\mu)\rho(\mu,\xi)\,d\mu= t(\lambda,\xi). \ee
It coincides with $\rho_{tot}(\lambda)$ at $\xi=\eta/2$. For our
goals it is enough to consider $-\zeta<\mathrm{Im} (\xi) <0$ for $-1 <
\Delta \le 1$ and $\eta< \mathrm{Re} (\xi) <0$ for $\Delta>1$. For zero
magnetic field, one has
\be{GFsolmasles}
\rho(\lambda,\xi) =\left\{
\begin{array}{cc}
{\dis\frac{i}{2\zeta\sinh\frac\pi\zeta(\lambda-\xi)},}&
{\dis\qquad |\Delta|<1,~\zeta=i\eta,}\non
{\dis\frac{i}{2\pi}\prod_{n=1}^\infty\left(\frac{1-q^{2n}}{1+q^{2n}}
\right)^2\frac{\vartheta_3(i(\lambda-\xi),q)}{\vartheta_4(i(\lambda-\xi),q)}
,}&{\dis\qquad \Delta>1,~q=e^\eta.}
\end{array}\right.
\ee
In the presence of the magnetic field, the equation \eq{GFinteqinh}
cannot be solved explicitely in terms of known elementary or special
functions. Some properties of $\rho(\lambda,\xi)$ can
nevertheless be established : in particular, it is not difficult to see that
$\rho(\lambda,\xi)$ has a simple pole at $\lambda=\xi$ with the
residue $2\pi i\Res\rho(\lambda,\xi)|_{\lambda=\xi}=-1$. This
property was used in \cite{KitMT00} for the computation of the
elementary blocks, and we shall also use it in Section \ref{ssc}.


\section{From quantum inverse scattering problem to correlation functions via algebraic Bethe ansatz  \label{CCF}}

In this section, we review the main steps of the method proposed in
\cite{KitMT99,KitMT00} for the computation of the correlation
functions in the framework of algebraic Bethe ansatz using the
solution of the quantum inverse scattering problem: we first recall
briefly how to obtain a multiple integral representation for the
elementary building blocks (\ref{genabcd00}) (see \cite{KitMT99,KitMT00}
for details), then
we discuss in this context the case of the spin-spin correlation
functions at lattice distance $m$.

\bigskip

The elementary blocks of the
correlation functions are defined as the normalized
expectation values of products of local matrices
$E^{\epsilon'_j,\epsilon_j}_j$ from site $j = 1$ to site $j = m$
with respect to some eigenstate of the transfer matrix
(for which the set of
spectral parameters $\lambda_k$ satisfy the Bethe equations):
\be{CCFev}
F_m(\{\epsilon_j,\epsilon'_j\})=\frac{\langle
0|\prod\limits_{k=1}^{N}C(\lambda_k)\ \biggl(\prod\limits_{j=1}^m
E^{\epsilon'_j,\epsilon_j}_j\biggr)\,
\prod\limits_{k=1}^{N}B(\lambda_k)|0\rangle} {\langle
0|\prod\limits_{k=1}^{N}C(\lambda_k)
\prod\limits_{k=1}^{N}B(\lambda_k)|0\rangle}.
\ee
For technical reasons, it is convenient to achieve the calculation in the
generic inhomogeneous case \eq{ABALinh}; it is easy at the end to
particularize the result to the homogeneous chain \eq{IHamXXZ}.

To compute the expectation values \eq{CCFev}, or more generally any kind of
correlation functions, one has first to express the
elementary local operators $E^{\epsilon'_j,\epsilon_j}_j$
(or equivalently the local spin operators) in terms of the entries of
the quantum monodromy matrix. Such
a representation is given by the solution of the quantum inverse
scattering problem \cite{KitMT99,MaiT00}:
\begin{thm}\cite{KitMT99,MaiT00}\label{pb-inv}
Let us consider the inhomogeneous $XXZ$ model (\ref{ABALinh}) with arbitrary
inhomogeneity parameters $\xi_k$, $1\le k \le M$.
The local spin operators at any site $j$ of the chain can be expressed in terms
of the elements of the quantum monodromy matrix as
\be{FCtab}
\begin{array}{l}
{\dis \sigma^-_j=\prod_{k=1}^{j-1} {\cal T}(\xi_k)\cdot B(\xi_j)
\cdot\prod_{k=1}^{j} {\cal T}^{-1}(\xi_k), }\non
{\dis
\sigma^+_j=\prod_{k=1}^{j-1} {\cal T}(\xi_k)\cdot C(\xi_j)
\cdot\prod_{k=1}^{j} {\cal T}^{-1}(\xi_k), }\non
{\dis
\sigma^z_j=\prod_{k=1}^{j-1} {\cal T}(\xi_k)\cdot
\bigl(A-D\bigr)(\xi_j)
\cdot\prod_{k=1}^{j} {\cal T}^{-1}(\xi_k). }
\end{array}
\ee
In particular, these formulas apply for the homogeneous model
(\ref{IHamXXZ}) where $\xi_k=\eta/2$, $1\le k \le M$.
\end{thm}
\begin{rem}\label{rem1}
The identity operator in the site $j$ can also be written in
a form similar to (\ref{FCtab}):
\be{FCidop}
{\bf 1}_j=\prod_{k=1}^{j-1} {\cal T}(\xi_k)\cdot
\bigl(A+D\bigr)(\xi_j) \cdot\prod_{k=1}^{j} {\cal T}^{-1}(\xi_k).
\ee
\end{rem}
\begin{rem}\label{rem2}
The resolution \eq{FCtab}--\eq{FCidop} of the quantum inverse scattering
problem can also be expressed in terms of the elementary matrices
$E^{\epsilon'_j,\epsilon_j}_j$ in the site $j$. In the homogeneous case, the
corresponding reconstruction formulas are given by \eq{ET}.
\end{rem}
From Theorem \ref{pb-inv} and Remark \ref{rem2}, one obtains
\begin{equation}
F_m(\{\epsilon_j,\epsilon'_j\})=\Phi_m(\{\lambda\})\frac
{\bra{\psi_g} T_{\epsilon_1,\epsilon'_1}(\xi_1)\dots
T_{\epsilon_m,\epsilon'_m}(\xi_m)\ket{\psi_g}}{\bra{\psi_g}\psi_g\rangle},
\label{genabcd}
\end{equation}
where $\Phi_m(\{\lambda\})$ is the ground state eigenvalue of the
corresponding product of the transfer matrices:
\be{phi}
\Phi_m(\{\lambda\})=\prod_{j=1}^m\prod_{a=1}^N\frac{\sinh(\lambda_a-\xi_j)}{\sinh(\lambda_a-\xi_j+\eta)}.
\ee

For the computation of these expectation values, one needs then to act
successively on the left with all the elements
$T_{\epsilon_m,\epsilon'_m}(\xi_m)$ of the monodromy matrix. One thus has to
use the expressions of the
action of the operators $A$, $B$, $C$, $D$ on an arbitrary state
$\langle\psi|=\prod_{j=1}^{N}C(\lambda_j)$. In the case of $A$ and
$D$, they are given by \cite{FadST79}:
\be{FCactA} \langle 0|\prod_{j=1}^{N}C(\lambda_j)
A(\lambda_{N+1})= \sum_{b=1}^{N+1}a(\lambda_b)
\frac{\prod\limits_{j=1}^{N}\sinh(\lambda_j-\lambda_b+\eta)}
{\prod\limits_{j=1\atop{j\ne b}}^{N+1}\sinh(\lambda_j-\lambda_b)}
\langle 0|\prod_{j=1\atop{j\ne b}}^{N+1}C(\lambda_j), \ee
\be{FCactD}
\langle 0|\prod_{j=1}^{N}C(\lambda_j) D(\lambda_{N+1})=
\sum_{a=1}^{N+1}d(\lambda_a)
\frac{\prod\limits_{j=1}^{N}\sinh(\lambda_a-\lambda_j+\eta)}
{\prod\limits_{j=1\atop{j\ne a}}^{N+1}\sinh(\lambda_a-\lambda_j)}
\langle 0|\prod_{j=1\atop{j\ne a}}^{N+1}C(\lambda_j).
\ee
The action of the operator $B$ is more complicated, and it is similar
to the successive action of $A$ and $D$:
\ba{FCactB}
&&{\dis\hspace{1mm}
\langle 0|\prod_{j=1}^{N}C(\lambda_j) B(\lambda_{N+1})=
\sum_{a=1}^{N+1}d(\lambda_a)
\frac{\prod\limits_{k=1}^{N}\sinh(\lambda_a-\lambda_k+\eta)}
{\prod\limits_{j=1\atop{k\ne
a}}^{N+1}\sinh(\lambda_a-\lambda_k)}}\nona{35}
&&{\dis\hspace{3mm}
\times\sum_{a'=1\atop{a'\ne a}}^{N+1}\frac{a(\lambda_{a'})}
{\sinh(\lambda_{N+1}-\lambda_{a'}+\eta)}
\frac{\prod\limits_{j=1\atop{j\ne a}}^{N+1}
\sinh(\lambda_j-\lambda_{a'}+\eta)}
{\prod\limits_{j=1\atop{j\ne a,a'}}^{N+1}
\sinh(\lambda_j-\lambda_{a'})}
\langle 0|\prod_{j=1\atop{j\ne a,a'}}^{N+1}C(\lambda_j).}
\ea
Finally, the action of the operator $C$ is free. Recall once more
that in the formulas \eq{FCactA}--\eq{FCactB} the parameters
$\{\lambda\}$ are arbitrary complex numbers (which are not necessarily
solutions of Bethe equations). In the above sums,
the terms containing $a(\lambda_{N+1})$ or $d(\lambda_{N+1})$ are
usually called direct terms, while the others are called indirect
terms.

The action of an arbitrary monomial
$T_{\epsilon_1,\epsilon'_1}(\xi_1)\dots
T_{\epsilon_m,\epsilon'_m}(\xi_m)$ on the state $\langle\psi|$ can
be obtained by applying recursively these formulas. This leads to a linear
combination of states,
\be{FCform} \langle 0|\prod_{k=1}^{N}C(\lambda_k)\cdot
T_{\epsilon_1,\epsilon'_1}(\xi_1)\dots
T_{\epsilon_m,\epsilon'_m}(\xi_m)= \sum_{i \in I} \alpha_i \bra{0}
\prod_{k \in K_i} C(\mu_k), \ee
with some (computable) coefficients $\alpha_i$. Here, sums and products are
taken over (multiple) sets $I$ and $K_i$, where the $K_i$, $i\in I$,
are subsets of $1, \dots, m+N$ with
$(\mu_1, \dots, \mu_{m+N}) = (\lambda_1, \dots, \lambda_N,\xi_1,
\dots, \xi_m)$.

\bigskip

Finally, to evaluate the expectation value
\eq{CCFev}, it remains to compute scalar products of the type
\be{CCFsp}
\langle 0|\prod_{j=1}^{N}C(\mu_j)
\prod_{j=1}^{N}B(\lambda_j)|0\rangle,
\ee
where $\prod_{j=1}^{N}B(\lambda_j)|0\rangle$ is an eigenstate of
the transfer matrix, while the parameters $\{\mu_j\}_{1\le j \le N}$
are arbitrary.
The result for \eq{CCFsp} is given by \cite{Sla89,Sla97} (see
\cite{KitMT99} for another proof):
\begin{prop}\label{prod-scal}
\cite{Sla89,Sla97,KitMT99}
The scalar product of a Bethe state with an arbitrary state of the form
\eq{ABAES} can be expressed in the following way:
\be{FCdet}
\langle 0|\prod_{j=1}^{N}C(\mu_j)
\prod_{j=1}^{N}B(\lambda_j)|0\rangle=
\frac{\prod\limits_{a,b=1}^{N}\sinh(\lambda_b-\mu_a+\eta)}
{\prod\limits_{a>b}^{N}\sinh(\mu_a-\mu_b)\sinh(\lambda_b-\lambda_a)}
{\det}_N\Psi'(\{\mu\}|\{\lambda\}),
\ee
for $\{\lambda_j\}_{1\le j \le N}$ solution of the Bethe equation, and for any
set of complex parameters $\{\mu_j\}_{1\le j \le N}$.
The $N\times N$ matrix $\Psi'(\{\mu\}|\{\lambda\})$ is defined by
\be{FCentri}
\Psi'_{jk}(\{\mu\}|\{\lambda\})=
t(\lambda_j,\mu_k)-d(\mu_k)t(\mu_k,\lambda_j)
\prod_{a=1}^{N}\frac{\sinh(\mu_k-\lambda_a+\eta)}
{\sinh(\mu_k-\lambda_a-\eta)}.
\ee
\end{prop}
Here and further ${\det}_N$ denotes the determinant of an $N\times
N$ matrix. Setting $\{\mu\}=\{\lambda\}$ in \eq{CCFsp}, one obtains
the square of the norm of the corresponding eigenstate
\cite{Kor82}:
\be{FCnorm}
\langle 0|\prod_{j=1}^{N}C(\lambda_j)
\prod_{j=1}^{N}B(\lambda_j)|0\rangle=\sinh^N\eta
\prod\limits_{a,b=1\atop{a\ne b}}^{N}
\frac{\sinh(\lambda_a-\lambda_b+\eta)}
{\sinh(\lambda_a-\lambda_b)}{\det}_N \Phi'(\{\lambda\}).
\ee
where
\be{GFPhi}
\Phi'_{jk}(\{\lambda\})
=\delta_{jk}\left[\frac{d'(\lambda_j)}{d(\lambda_j)}
-\sum_{a=1}^{N} K(\lambda_j-\lambda_a)\right]+K(\lambda_j-\lambda_k),
\ee
with $K(\lambda)$ given in \eq{GFKkern}. Note that both matrices
$\Psi'$ and $\Phi'$ can be written in the form of Jacobians
\cite{KitMT99,Kor82}. We also would like to point out that the
entries of the matrix $\Psi'$ are linear combinations of
$t(\lambda_j,\mu_k)$ and $t(\mu_k,\lambda_j)$. In the next section
we also deal with determinants of matrices possessing similar
structure. In fact all these determinants are various deformations
of $\det t(\lambda_j,\mu_k)$ describing the partition function of
the six-vertex model with domain wall boundary conditions
\cite{Ize87}. An explanation of this deformation was given in
\cite{KitMT99}.

Using the formulas \eq{FCtab}--\eq{GFPhi}, one can compute the
normalized average value  \eq{CCFev} on the finite lattice. The
remaining step is to proceed to the thermodynamic limit. Observe
that the actions \eq{FCactA}--\eq{FCactB} produce  sums with
respect to parameters $\{\lambda\}$. The successive action of
several $E^{\epsilon'_j,\epsilon_j}_j$ (i.e. successive action of
the entries of the monodromy matrix) gives multiple sums. Let the
parameters $\lambda_j$ describe the ground state; then, in the
thermodynamic limit, each of these sums turns into the integral
\be{CCFsumint}
\frac1M\sum_{\{\lambda\}}f(\lambda)
\longrightarrow\int_{C}
f(\lambda)\rho_{tot}(\lambda)\,d\lambda,
\ee
where $\rho_{tot}(\lambda)$ is the solution of Lieb equation
\eq{GFLiebeq}. It was shown also in \cite{KitMT00} that in the
thermodynamic limit the ratio of the determinants of $\Psi'$ and
$\Phi'$ can be evaluated as a determinant  of inhomogeneous
densities \eq{GFinteqinh}. Namely, if in \eq{FCdet} we have
$\{\mu\}=\{\xi_1,\dots,\xi_n\}\cup
\{\lambda_{n+1},\dots,\lambda_{N}\}$,
\be{GFrat2det}
\frac{{\det}_N \Psi'}{{\det}_N\Phi'}
=\prod_{a=1}^n(M\rho_{tot}(\lambda_a))^{-1}
{\det}_{n}\rho(\lambda_j,\xi_k).
\ee
In \cite{KitMT00}, it has also been shown that, when acting with the
monodromy matrix elements, one can get rid of all direct type
terms in the thermodynamic limit: the procedure is just to shift
properly the integration contour of the corresponding $\lambda$
variables. In this way, the expectation value of
$T_{\epsilon_1,\epsilon'_1}(\xi_1)\dots
T_{\epsilon_m,\epsilon'_m}(\xi_m)$ in the ground state is
represented as a multiple integral in which the number of integrals
coincides with the number of operators in the product. This
quantity is called an elementary block.

\bigskip

This method gives generic answers for the computation of the
expectation values of the monomials of the type
$T_{\epsilon_1,\epsilon'_1}(\xi_1)\dots
T_{\epsilon_m,\epsilon'_m}(\xi_m)$ corresponding to elementary
blocks of the correlation functions \cite{KitMT99,KitMT00}. The
problem of the evaluation of the spin-spin correlation functions
is more involved. Let us consider, for example, the correlation
function $\langle\sigma_1^+\sigma_{m+1}^-\rangle$.
From the solution of the inverse scattering problem (Theorem \ref{pb-inv}),
one obtains the identity
\be{CCF+-} \langle\psi|\,\sigma_1^+\,\sigma_{m+1}^-|\psi\rangle \equiv
\langle\psi|\,C(\xi_1)\cdot\prod_{a=2}^{m}(A+D)(\xi_a)\cdot B(\xi_{m+1})
\cdot
\prod_{b=1}^{m+1}(A+D)^{-1}(\xi_b)\,|\psi\rangle. \ee
Here $|\psi\rangle$ is an eigenstate of the transfer matrix $(A+D)$,
and therefore it is straightforward to act on the right with
$\prod_{b=1}^{m+1}(A+D)^{-1}(\xi_b)$.
However, the determination of the action of $\prod_{a=2}^{m}(A+D)(\xi_a)$
is more involved: clearly after acting with $C(\xi_1)$ on $\langle\psi|$
(or equivalently with $B(\xi_{m+1})$ on $|\psi\rangle$), one obtains
a sum of states which are no longer Bethe states; therefore the
multiple action of $(A+D)$ on these states is not simple. In fact,
in the framework of the above approach, the product
$\prod_{a=2}^{m}(A+D)(\xi_a)$ would be computed  as a sum of
$2^{m-1}$ monomials, which eventually leads us to the sum of
$2^{m-1}$ elementary blocks. As already discussed in
Introduction, this form of the result is not suitable, in
particular at large distance $m$. Therefore, to obtain manageable
(re-summed) expressions for spin-spin correlation functions, it is
essential to obtain an alternative and compact evaluation of the
multiple action of the transfer matrix on arbitrary states. This
is the subject of the next section.


\section{The multiple action of the transfer matrix
$(A+D)(x)$\label{AP}}

The new results obtained in this section play a central role in
the computation of spin-spin correlation functions at distance
$m$: we evaluate here the multiple action of the
transfer matrix $A+D$ on an arbitrary state, which enables us to solve the
problem mentioned at the end of the previous section.

More precisely, let us consider the product
\be{APprod}
\prod_{a=1}^m\left(A+e^\beta D\right)(x_a),
\ee
where $x_1,\dots,x_m$ and $\beta$ are arbitrary complex. Our aim
is to find a compact formula for the action of this operator on
a state $\langle\psi|=\langle 0|\prod_{j=1}^NC(\lambda_j)$, where
parameters $\lambda_j$ are also arbitrary numbers. For simplicity,
we first  consider the case $m\le N$.

Due to \eq{FCactA}, \eq{FCactD}, the action of single operator
$\left(A+e^\beta
 D\right)(x)$ on the state $\langle\psi|$ can be written in
the form:
\be{APactsing}
\langle 0|\prod_{j=1}^NC(\lambda_j)
\left(
A+e^\beta D\right)(x)=
\Lambda\langle 0|\prod_{j=1}^NC(\lambda_j)
+\sum_{k=1}^N \Lambda_k
\langle 0|\prod_{j=1\atop{j\ne k}}^NC(\lambda_j)
\cdot C(x).
\ee
Here
\be{APdirsing} \Lambda= a(x)\prod_{j=1}^Nf(\lambda_j,x)+ e^\beta
d(x)\prod_{j=1}^Nf(x,\lambda_j), \ee
and
\be{APundirsing}
\Lambda_k=
a(\lambda_k)\frac{\sinh\eta}{\sinh(x-\lambda_k)}
\prod_{j=1\atop{j\ne k}}^Nf(\lambda_j,\lambda_k)+
e^\beta d(\lambda_k)\frac{\sinh\eta}{\sinh(\lambda_k-x)}
\prod_{j=1\atop{j\ne k}}^Nf(\lambda_k,\lambda_j).
\ee
Recall that $a(\lambda)\equiv 1$; however, up to the end of this
section, we do not use this property, nor the explicit form of
$d(\lambda)$. All our derivations are based on the use of the
equations \eq{FCactA}, \eq{FCactD}, which in turn are direct
corollary of the intertwining relation \eq{ABARTT}. Thus, the
result \eq{APcompl-act}, \eq{APRn} below for the action of the operator
\eq{APprod} on an arbitrary state is valid for any model
with an $R$-matrix of the form \eq{ABAR}.

Let us denote the first and the second type of action in the r.h.s. of
\eq{APactsing} by `direct' and `indirect' actions respectively.
The corresponding coefficients \eq{APdirsing} and \eq{APundirsing} are
called direct and indirect terms. The direct action
of $\left(A+e^\beta D\right)(x)$ preserves the state
$\langle\psi|$, as if this state was an eigenstate of the operator.
The indirect
action leads to the remaining terms, where the resulting states
differ from the original one: in these states, the argument $x$ of the
operator $\left(A+e^\beta D\right)(x)$  becomes the argument of one of the
operators $C$.

One can similarly define direct and indirect action for the operator
\eq{APprod}. Namely,  the direct action of this operator
corresponds to the term where the resulting state coincides with
the original one, which gives
\be{APdir-act}
\langle\psi|\left.\prod_{a=1}^m\left(A+e^\beta D\right)(x_a)
\right|_{(dir.)}=
\prod_{a=1}^m\biggl\{
a(x_a)\prod_{j=1}^Nf(\lambda_j,x_a)+
e^\beta d(x_a)\prod_{j=1}^Nf(x_a,\lambda_j)
\biggr\}\langle\psi|.
\ee
All the remaining terms result from the indirect action.
For $m \le N$, a particular kind of indirect terms corresponds to the case
where the whole set of arguments
$\{x\}$ of the operators $\left(A+e^\beta D\right)(x_a)$ enters the
resulting states. The corresponding action is called `completely
indirect'. It can be written in the form
\ba{APundir-act}
&&{\dis\hspace{-11mm}
\langle\psi|
\left.\prod_{a=1}^m\left(A+e^\beta D\right)(x_a)
\right|_{(c.-ind.)}}\non
&&{\dis\hspace{1mm}
=\sum_{\{\lambda\}=\{\lambda_{\alpha_+}\}\cup\{\lambda_{\alpha_-}\}
\atop{|\alpha_+|=m}}
S_m(\{x\}|\{\lambda_{\alpha_+}\}|\{\lambda_{\alpha_-}\})\
\langle 0| \prod_{a=1}^m C(x_a)
\prod_{b\in\alpha_-}C(\lambda_b).}
\ea
Here the set of the parameters $\{\lambda\}$ is divided into two
subsets:
$\{\lambda\}=\{\lambda_{\alpha_+}\}\cup\{\lambda_{\alpha_-}\}$.
The subset $\{\lambda_{\alpha_+}\}$ is replaced with $\{x\}$ in
the resulting states, and therefore  the number of elements in
this subset is equal $m$ (we have indicated this fact in
\eq{APundir-act} by $|\alpha_+|=m$). The subset
$\{\lambda_{\alpha_-}\}$ remains in the resulting states. The sum
in \eq{APundir-act} is taken with respect to all such partitions
of the parameters $\{\lambda\}$. Following the tradition, we call
the corresponding factor
$S_m(\{x\}|\{\lambda_{\alpha_+}\}|\{\lambda_{\alpha_-}\})$ the
highest coefficient.
\begin{lemma}\label{+coeff}
The highest coefficient $S_m$ in \eq{APundir-act} can be expressed in the
following form:
\be{APSm}
S_m(\{x\}|\{\lambda_{\alpha_+}\}|\{\lambda_{\alpha_-}\}) =
\frac{\prod\limits_{b\in\alpha_+}\prod\limits_{a=1}^m
\sinh(x_a-\lambda_b+\eta)}
{\prod\limits_{a>b\atop{a,b\in\alpha_+}}
\sinh(\lambda_a-\lambda_b)
\prod\limits_{a<b}^m
\sinh(x_a-x_b)}\cdot
\det_{j\in\alpha_+\atop{k=1,\dots,m}}M_{jk}~,
\ee
where the $m\times m$ matrix $M$ is given by
\be{APMjk}
\hspace{-2mm}
M_{jk}=a(\lambda_j)t(x_k,\lambda_j)
\prod_{a\in\alpha_-}f(\lambda_a,\lambda_j)
-e^\beta d(\lambda_j)t(\lambda_j,x_k)
\prod_{a\in\alpha_-}f(\lambda_j,\lambda_a)
\prod_{b=1}^m\frac{\sinh(\lambda_j-x_b+\eta)}
{\sinh(\lambda_j-x_b-\eta)},
\ee
\end{lemma}
The proof is given in Appendix \ref{HC}.

Thus, we have described two limiting cases in the action of the
operator \eq{APprod}: none of the parameters $\{x\}$ enter the
final state, or all of them enter the final state. The complete
action of the operator \eq{APprod} contains also all the intermediate
cases, when only a subset of the parameters $\{x\}$ enters the
final state.

\begin{prop}\label{act-t}
Let $p = \min(m,N)$. The action of $\prod_{a=1}^m\left(A+e^\beta
D\right)(x_a)$ on a general state $\langle\psi|=\langle
0|\prod_{j=1}^NC(\lambda_j)$, for any sets of complex parameters
$\{x_a\}_{1\le a \le  m}$ and $\{\lambda_j\}_{1\le j\le N}$,
can be written as
\ba{APcompl-act} &&{\dis\hspace{-2mm}
\langle\psi|\prod_{a=1}^m\left(A+e^\beta D\right)(x_a)}\non
&&{\dis\hspace{1mm}
=\sum_{n=0}^p\sum_{\{\lambda\}=\{\lambda_{\alpha_+}\}\cup\{\lambda_{\alpha_-}\}
\atop{\{x\}=\{x_{\gamma_+}\}\cup\{x_{\gamma_-}\}
\atop{|\alpha_+|=|\gamma_+|=n}}}
R_n(\{x_{\gamma_+}\}|\{x_{\gamma_-}\}|\{\lambda_{\alpha_+}\}|\{\lambda_{\alpha_-}\})
\langle 0 |\prod_{a\in\gamma_+} C(x_a)
\prod_{b\in\alpha_-}C(\lambda_b).} \ea
The coefficient $R_n$ in \eq{APcompl-act} is given by
\ba{APRn}
&&{\dis\hspace{-7mm}
R_n(\{x_{\gamma_+}\}|\{x_{\gamma_-}\}|\{\lambda_{\alpha_+}\}|\{\lambda_{\alpha_-}\})
=S_n(\{x_{\gamma_+}\}|\{\lambda_{\alpha_+}\}|\{\lambda_{\alpha_-}\})}\non
&&{\dis\hspace{-13mm}
\times\prod_{a\in\gamma_-}\biggl\{
a(x_a)\prod_{b\in\gamma_+}f(x_b,x_a)
\prod_{b\in\alpha_-}f(\lambda_b,x_a)+
e^\beta d(x_a)
\prod_{b\in\gamma_+}f(x_a,x_b)
\prod_{b\in\alpha_-}f(x_a,\lambda_b)
\biggr\}.}
\ea
\end{prop}
{\sl Proof.}~~ Let us start with the case $p = m$. Note first
that, in this formula, we deal with two partitions of the
sets $\{\lambda\}$ and $\{x\}$: the subset
$\{\lambda_{\alpha_+}\}$ is replaced with the parameters
$\{x_{\gamma_+}\}$ in the final state, and the number of elements
in these subsets is equal to $n$; the remaining variables
$\{x_{\gamma_-}\}$  do not enter the final state. Once again, the sum is
taken with respect to all possible partitions of such type. The
limiting cases $n=0$ and $n=m$ correspond to the direct and
completely indirect actions respectively.

One can now use standard considerations of algebraic Bethe ansatz.
Namely, let us fix a certain partition in \eq{APcompl-act} and try to obtain
only the state $\langle 0 |\prod_{a\in\gamma_+}
C(x_a)\prod_{b\in\alpha_-}C(\lambda_b)$, applying the operator \eq{APprod} to
the original state $\langle\psi|$. Due to their commutativity, one can
re-order the operators in the product \eq{APprod} as
\be{APreoder}
\prod_{a=1}^m\left(A+e^\beta D\right)(x_a)=
\prod_{a\in\gamma_+}\left(A+e^\beta D\right)(x_a)\cdot
\prod_{a\in\gamma_-}\left(A+e^\beta D\right)(x_a).
\ee
Then it is easy to see that the action of the first group of operators
must be completely indirect, otherwise one of the parameters $x_a$,
$a\in \gamma_+$, is missing in the set of arguments of the final state.
Moreover, the subset of parameters replaced in the
original state $\langle\psi|$ must be exactly equal to
$\{\lambda_{\alpha_+}\}$:
otherwise, at least one of the elements of $\{\lambda_{\alpha_+}\}$ or
of $\{x_{\gamma_-}\}$ belongs to the final state.
Therefore, the action of the first product of operators contributes as
\ba{AP1step}
&&{\dis\hspace{-11mm}
\langle\psi|\prod_{a\in\gamma_+}\left(A+e^\beta D\right)(x_a)
\prod_{a\in\gamma_-}\left(A+e^\beta D\right)(x_a)
\longrightarrow}\non
&&{\dis\hspace{1mm}
S_n(\{x_{\gamma_+}\}|\{\lambda_{\alpha_+}\}|\{\lambda_{\alpha_-}\})\
\langle 0| \prod_{a\in\gamma_+} C(x_a)
\prod_{b\in\alpha_-}C(\lambda_b)
\prod_{a\in\gamma_-}\left(A+e^\beta D\right)(x_a).}
\ea
In its turn, the action of the remaining group of operators
$\left(A+e^\beta D\right)(x_a)$ in \eq{AP1step} must be direct,
otherwise one of the elements of $\{x_{\gamma_-}\}$ should
appear in the final state.
Using \eq{APdir-act} we immediately arrive at \eq{APRn}.

This proof obviously extends immediately to the case $m \ge N$.
We just have in this case additional  direct type actions, and
hence no completely indirect term. Thus, the action of the operator
\eq{APprod} on the state $\langle\psi|$ is given by
\eq{APcompl-act} with the coefficients $R_n$ defined in
\eq{APRn}.
\qed


For our purposes, we will need to specify the arguments $\{x\}$ of
the operator \eq{APprod}. In the inhomogeneous model, we should set
them equal to the inhomogeneity  parameters: $x_j=\xi_j$.
Then, since $d(\xi_j)=0$, the
part of $R_n$ corresponding to the direct action simplifies
(the operator $D(\xi_j)$ has no direct action). Thus,
the equation \eq{APcompl-act} takes the form ($p = \min(m,N)$):
\ba{APactxi} &&{\dis\hspace{-2mm}
\langle\psi|\prod_{a=1}^m\left(A+e^\beta D\right)(\xi_a)
=\sum_{n=0}^p\hspace{4mm}
\sum_{\{\lambda\}=\{\lambda_{\alpha_+}\}\cup\{\lambda_{\alpha_-}\}
\atop{\{\xi\}=\{\xi_{\gamma_+}\}\cup\{\xi_{\gamma_-}\}
\atop{|\alpha_+|=|\gamma_+|=n}}}
S_n(\{\xi_{\gamma_+}\}|\{\lambda_{\alpha_+}\}|\{\lambda_{\alpha_-}\})
}\non
&&{\dis\hspace{11mm}
\times
\prod_{a\in\gamma_-}\biggl\{
a(\xi_a)\prod_{b\in\gamma_+}f(\xi_b,\xi_a)
\prod_{b\in\alpha_-}f(\lambda_b,\xi_a)\biggr\}\,
\langle 0 |\prod_{a\in\gamma_+} C(\xi_a)
\prod_{b\in\alpha_-}C(\lambda_b).}
\ea

Finally, one can also consider the particular case when
$\langle\psi|$ is an eigenstate of the transfer matrix. Since the
parameters $\{\lambda\}$ satisfy the system of Bethe equations
\eq{ABABE}, one obtains:
\begin{cor}
If $\langle\psi|$ is an eigenstate of the transfer matrix,
that is if the parameters $\{\lambda\}$ satisfy the system of Bethe
equations \eq{ABABE}, the multiple action of $(A+e^\beta
D)(\xi_j)$ is given by
\ba{APactxies} &&{\dis\hspace{-2mm}
\langle\psi|\prod_{a=1}^m\left(A+e^\beta D\right)(\xi_a)
=\sum_{n=0}^p\hspace{4mm}
\sum_{\{\lambda\}=\{\lambda_{\alpha_+}\}\cup\{\lambda_{\alpha_-}\}
\atop{\{\xi\}=\{\xi_{\gamma_+}\}\cup\{\xi_{\gamma_-}\}
\atop{|\alpha_+|=|\gamma_+|=n}}} \tilde
S_n(\{\xi_{\gamma_+}\}|\{\lambda_{\alpha_+}\}) }\non
&&{\dis\hspace{-6mm}
\times
\prod_{a\in\gamma_-}\prod_{b\in\gamma_+}f(\xi_b,\xi_a)
\prod_{a\in\gamma_-}\prod_{b\in\alpha_-}f(\lambda_b,\xi_a)
\prod_{a\in\alpha_-}\prod_{b\in\alpha_+}f(\lambda_a,\lambda_b)
\langle 0 |\prod_{a\in\gamma_+} C(\xi_a)
\prod_{b\in\alpha_-}C(\lambda_b).}
\ea
Here $p = \min(m,N)$, and
\be{APtiSn}
\tilde S_n(\xi_1,\dots\xi_n|\lambda_1\dots,\lambda_n) =
\frac{\prod\limits_{a=1}^n\prod\limits_{b=1}^n
\sinh(\xi_a-\lambda_b+\eta)}
{\prod\limits_{a>b}^n
\sinh(\lambda_a-\lambda_b)
\prod\limits_{a<b}^n
\sinh(\xi_a-\xi_b)}\cdot
\det\tilde M_{jk}~,
\ee
where the $n\times n$ matrix $\tilde M_{jk}$ is
\be{APtiMjk}
\hspace{-2mm}
\tilde M_{jk}=t(\xi_k,\lambda_j)
+e^\beta t(\lambda_j,\xi_k)
\prod_{a=1}^n\frac{\sinh(\lambda_a-\lambda_j+\eta)}
{\sinh(\lambda_j-\lambda_a+\eta)}
\prod_{a=1}^n\frac{\sinh(\lambda_j-\xi_a+\eta)}
{\sinh(\xi_a-\lambda_j+\eta)}.
\ee
\end{cor}
Here we have set $a(\lambda)=1$. The remarkable property of
$\tilde S_n$ is that $\tilde S_n=\delta_{n,0}$ at $\beta=0$ (see
Appendix \ref{Sn}). Thus, in this case, all the terms in
\eq{APactxies} with $n\ge1$ vanish.  The subsets
$\{\lambda_{\alpha_+}\}$ and $\{\xi_{\gamma_+}\}$ are empty, and
we arrive at
\be{APactA+Des}
\langle\psi|\prod_{a=1}^m\left(A+D\right)(\xi_a)=
\prod_{a=1}^m\prod_{b=1}^Nf(\lambda_b,\xi_a)
\langle\psi|,
\ee
as it should be for the eigenstates of the transfer matrix.


\section{Generating functional for the $\sigma^z$ correlation function \label{GF}}

In the present and in the next sections, we use the above results
to compute the spin-spin correlation functions
$\langle\sigma_1^\alpha\sigma_{m+1}^\beta\rangle$.
As we are in the framework of algebraic Bethe ansatz, we begin with the
evaluation of normalized expectation values on the finite lattice
with respect to an arbitrary eigenstate $|\psi\rangle=
\prod_{j=1}^{N}B(\lambda_j)|0\rangle$. In this case, the
relationship between $m$ and $N$ is not fixed. However, for the
ground state in the thermodynamic limit, $N$ tends to infinity while $m$
remains finite. Therefore it is clear that we can consider the
case $m<N$ only.

\bigskip

One of the simplest applications of the formulas obtained in the
previous section is the computation of the generating functional
for the correlation function of the third components of spins
$\langle\sigma_1^z\sigma_{m+1}^z\rangle$. Following
\cite{IzeK85,ColIKT93}, we define an operator $Q_{1,m}$ as
\be{GFdefQ}
Q_{1,m}=\frac12\sum_{k=1}^{m}(1-\sigma_k^z).
\ee
The generating functional is equal to the expectation value
$\langle\exp(\beta Q_{1,m})\rangle$,  where $\beta$ is some
complex number. In particular, taking the second derivative of
this quantity with respect to $\beta$ at $\beta=0$, we obtain
$\langle Q_{1,m}^2\rangle$. Taking then the second lattice
derivative we have:
\be{GFlatder}
\frac12\langle(1-\sigma_1^z)(1-\sigma_{m+1}^z)\rangle=
\langle Q_{1,m+1}^2\rangle-\langle Q_{1,m}^2\rangle-
\langle Q_{2,m+1}^2\rangle+\langle Q_{2,m}^2\rangle.
\ee
Thus, the two-point correlation function of the third components
of the local spins can be easily extracted from the expectation
value of the operator $\exp(\beta Q_{1,m})$ in the homogeneous
limit.
\begin{prop}\label{Q}
The correlation function $\langle\sigma_1^z\sigma_{m+1}^z\rangle$
in the homogeneous model is given by
\be{Ogenfun} \langle\sigma_1^z\sigma_{m+1}^z\rangle= \left(2{\cal
D}^2_m\left.\frac{
\partial^2}{\partial\beta^2}
-4{\cal D}_m\frac{
\partial}{\partial\beta}+1\right)
\langle\exp(\beta Q_{1,m})\rangle \right|_{\beta=0}, \ee
where the symbols ${\cal D}_m$ and ${\cal D}^2_m$ mean the first
and the second lattice derivative respectively:
\be{Olatder} {\cal D}_mf(m)\equiv f(m+1)-f(m),\qquad {\cal
D}_m^2f(m)\equiv f(m+1)+f(m-1)-2f(m). \ee

The expression of the ground state expectation value of the operator
$\exp(\beta Q_{1,m})$ in the inhomogeneous case is
\ba{GFebQm+1}
&&{\dis\hspace{1mm}
\langle\exp(\beta
Q_{1,m})\rangle=\sum_{n=0}^{m} \frac1{(n!)^2}\oint\limits_\Gamma
\prod_{j=1}^n\frac{dz_j}{2\pi i} \int_C d^n\lambda
\prod_{b=1}^n\prod_{a=1}^{m} \frac{f(z_b,\xi_a)} {
f(\lambda_b,\xi_a)} }\non
&&{\dis\hspace{21mm} \times W_n(\{\lambda\},\{z\})\cdot
{\det}_{n}\Bigl[\tilde M_{jk}(\{\lambda\}|\{z\})\Bigr]
\cdot{\det}_{n}\Bigl[\rho(\lambda_j,z_k)\Bigr].} \ea
Here
\be{GFWn}
W_n(\{\lambda\},\{z\})= \prod_{a=1}^n\prod_{b=1}^n\frac{
\sinh(\lambda_a-z_b+\eta)\sinh(z_b-\lambda_a+\eta)}
{\sinh(\lambda_a-\lambda_b+\eta)\sinh(z_a-z_b+\eta)}, \ee
and
\be{GFtiMjk}
\hspace{-2mm} \tilde M_{jk}(\{\lambda\}|\{z\})
=t(z_k,\lambda_j)+e^\beta t(\lambda_j,z_k)
\prod_{a=1}^n\frac{\sinh(\lambda_a-\lambda_j+\eta)}
{\sinh(\lambda_j-\lambda_a+\eta)} \frac{\sinh(\lambda_j-z_a+\eta)}
{\sinh(z_a-\lambda_j+\eta)}. \ee
The contour $C$, as in the Lieb equation \eq{GFLiebeq}, depends on the regime
and on the value of the magnetic field, and $\Gamma$ surrounds only the
singularities at the inhomogeneity parameters $\xi_k$, $1\le k\le m$.

In the homogeneous limit $\xi_j=\eta/2$, the generating functional
of the two-point function can be written as
\ba{GFebQm+1hom}
&&{\dis\hspace{1mm}
\langle\exp(\beta
Q_{1,m})\rangle= \sum_{n=0}^{m} \frac1{(n!)^2}\oint\limits_\Gamma
\prod_{j=1}^n\frac{dz_j}{2\pi i} \int_C d^n\lambda \prod_{a=1}^n
\left(\frac{\sinh(z_a+\frac\eta2)\sinh(\lambda_a-\frac\eta2)}
{\sinh(z_a-\frac\eta2)\sinh(\lambda_a+\frac\eta2)}\right)^{m}
}\non
&&{\dis\hspace{21mm} \times  W_n(\{\lambda\},\{z\})\cdot
{\det}_{n}\Bigl[\tilde M_{jk}(\{\lambda\}|\{z\})\Bigr]
\cdot{\det}_{n}\Bigl[\rho(\lambda_j,z_k)\Bigr],} \ea
where $\Gamma$ surrounds the point $\eta/2$.
\end{prop}
{\sl Proof.}~~The operator $\exp(\beta Q_{1,m})$ has a very simple
representation in terms of the entries of the monodromy matrix.
Due to \eq{FCtab} we have
\be{GFexpbQ} \exp(\beta Q_{1,m})=\prod_{a=1}^{m}\left(A+e^\beta
D\right)(\xi_a)\cdot \prod_{b=1}^{m}\left(A+D\right)^{-1}(\xi_b). \ee
Hence, we can directly apply the equation \eq{APactxies} for the
computation of $\langle\exp(\beta Q_{1,m})\rangle$. We begin with
the derivation of the normalized expectation value of $\exp(\beta
Q_{1,m})$ on the finite lattice with respect to an arbitrary
eigenstate of the transfer matrix,
$|\psi\rangle=\prod_{j=1}^NB(\lambda_j) |0\rangle$:
\be{GFexpbQev}
\frac{\langle\psi|\exp(\beta Q_{1,m})|\psi\rangle}
{\langle\psi|\psi\rangle}
=\frac{\langle\psi|\prod\limits_{a=1}^{m}
\left(A+e^\beta D\right)(\xi_a)
\prod\limits_{b=1}^{m}(A+D)^{-1}(\xi_b)|\psi\rangle}
{\langle\psi|\psi\rangle}.
\ee
The action of the operators $(A+D)^{-1}(\xi_b)$ to the right is trivial:
\be{GFactright}
\prod\limits_{b=1}^{m}(A+D)^{-1}(\xi_b)|\psi\rangle=
\prod\limits_{b=1}^{m} \prod_{a=1}^N f^{-1}(\lambda_a,\xi_b)|\psi\rangle.
\ee
The action of the operators $\left(A+e^\beta D\right)(\xi_a)$
to the left is given by
\eq{APactxies}--\eq{APtiMjk}. We obtain
\ba{GFactxies}
&&{\dis\hspace{-2mm}
\frac{\langle\psi|\exp(\beta Q_{1,m})|\psi\rangle}
{\langle\psi|\psi\rangle}
=\prod_{b=1}^{m} \prod_{a=1}^N f^{-1}(\lambda_a,\xi_b)
\sum_{n=0}^m\hspace{4mm}
\sum_{\{\lambda\}=\{\lambda_{\alpha_+}\}\cup\{\lambda_{\alpha_-}\}
\atop{\{\xi\}=\{\xi_{\gamma_+}\}\cup\{\xi_{\gamma_-}\}
\atop{|\alpha_+|=|\gamma_+|=n}}}
\tilde S_n(\{\xi_{\gamma_+}\}|\{\lambda_{\alpha_+}\})
}\non
&&{\dis\hspace{12mm}
\times
\prod_{a\in\gamma_-}\prod_{b\in\gamma_+}f(\xi_b,\xi_a)
\prod_{a\in\gamma_-}\prod_{b\in\alpha_-}f(\lambda_b,\xi_a)
\prod_{a\in\alpha_-}\prod_{b\in\alpha_+}f(\lambda_a,\lambda_b)
}\non
&&{\dis\hspace{32mm}
\times
\frac{\langle 0 |\prod\limits_{a\in\gamma_+} C(\xi_a)
\prod\limits_{b\in\alpha_-}C(\lambda_b)
\prod\limits_{j=1}^N B(\lambda_j)|0\rangle}
{\langle0|\prod\limits_{j=1}^N C(\lambda_j)
\prod\limits_{j=1}^N B(\lambda_j)|0\rangle}.}
\ea
In fact, for the remaining part of the calculations, we can use the formulas
of Section \ref{CCF}.
Using \eq{FCdet}--\eq{GFPhi} we find the ratio of the scalar products in
the r.h.s.
 of \eq{GFactxies} and finally arrive at
\ba{GFebQfl}
&&{\dis\hspace{-14mm}
\frac{\langle\psi|\exp(\beta Q_{1,m})|\psi\rangle}
{\langle\psi|\psi\rangle}
=
\sum_{n=0}^m\hspace{4mm}
\sum_{\{\lambda\}=\{\lambda_{\alpha_+}\}\cup\{\lambda_{\alpha_-}\}
\atop{\{\xi\}=\{\xi_{\gamma_+}\}\cup\{\xi_{\gamma_-}\}
\atop{|\alpha_+|=|\gamma_+|=n}}}
\prod_{b=1}^{m} \prod_{a\in\alpha_+} f^{-1}(\lambda_a,\xi_b)
\prod_{a\in\gamma_-}\prod_{b\in\gamma_+}f(\xi_b,\xi_a)
}\non
&&{\dis\hspace{-14mm}
\times
\frac{\prod\limits_{a\in\gamma_+}\prod\limits_{b\in\alpha_+}
\sinh(\xi_a-\lambda_b+\eta)\sinh(\lambda_b-\xi_a+\eta)}
{\prod\limits_{a,b\in\gamma_+\atop{a\ne b}}
\sinh(\xi_a-\xi_b)\prod\limits_{a,b\in\alpha_+}
\sinh(\lambda_a-\lambda_b+\eta)}\cdot
\det_{j\in\alpha_+\atop{k\in\gamma_+}}\tilde M_{jk}\cdot
\frac{{\det}_N \Psi'(\{\xi_{\gamma_+}\}\cup\{\lambda_{\alpha_-}\}
|\{\lambda\})
}{{\det}_N\Phi'(\{\lambda\}) }.}
\ea
The representation \eq{GFebQfl} gives us the expectation value of the
operator $\exp(\beta Q_{1,m})$ on the finite lattice.

Up to this stage, our derivation was purely algebraic. Now we
should proceed to the thermodynamic limit.
In spite of the fact that this limit
strongly depends on the phase of the model, one can present the
final result in a quite general form.
First, using \eq{GFrat2det},
we have in the thermodynamic limit
\be{GFrat2det1}
\frac{{\det}_N \Psi'(\{\xi_{\gamma_+}\}\cup\{\lambda_{\alpha_-}\}
|\{\lambda\})
}{{\det}_N\Phi'(\{\lambda\}) }
=\prod_{a\in\alpha_+}(M\rho_{tot}(\lambda_a))^{-1}
{\det}_{n}\rho(\lambda_j,\xi_k),
\ee
where $\lambda_j\in\{\lambda_{\alpha_+}\}$ and $\xi_k\in\{\xi_{\gamma_+}\}$.
Second, we should get rid of the sum with respect to
partitions in \eq{GFebQfl}, replacing them with integrals.

Consider first the partitions of the set $\{\xi\}$.
Let ${\cal F}(z_1,\dots,z_n)$ be a symmetric function of $n$ variables
$z_k$, analytical with respect to each argument in the vicinities of
$\{\xi\}$. Then
\ba{GFcontint}
&&{\dis\hspace{2mm}
\sum_{ \{\xi\}=\{\xi_{\gamma_-}\}\cup\{\xi_{\gamma_+}\}
\atop{ |\gamma_+|=n }}
\prod_{a\in\gamma_-}\prod_{b\in\gamma_+}
f(\xi_b,\xi_a)\cdot {\cal F}(\{\xi_{\gamma_+}\})}\non
&&{\dis\hspace{22mm}
=\frac1{n!}\int\limits_\Gamma \prod_{j=1}^n\frac{dz}{2\pi i}
\prod_{a=1}^n\prod_{b=1}^{m} f(z_a,\xi_b)
\frac{\prod\limits_{a=1}^n\prod\limits_{b=1\atop{b\ne a}}^n
\sinh(z_a-z_b)}
{\prod\limits_{a=1}^n\prod\limits_{b=1}^n
\sinh(z_a-z_b+\eta)}{\cal F}(\{z\}).}
\ea
Here the contour $\Gamma$ surrounds the points $\xi_1,\dots,
\xi_{m}$ and does not contain any other singularities of the integrand.
In particular, when all $\xi_j\to\eta/2$ (homogeneous limit), one
can chose $\Gamma$ as an enough small circle around $\eta/2$. Observe also
that one can easily take the homogeneous limit in the r.h.s. of
\eq{GFcontint}, while the
existence of such limit in the l.h.s. is not so obvious.
Thus, we can replace the sum with respect to the partitions of the set $\{\xi\}$
with the set of contour integrals. In fact, we could do this already on the
finite lattice.

As for the sum of the partitions of the set $\{\lambda\}$, we
essentially use the properties of the thermodynamic limit. Let now
${\cal F}(\lambda_1,\dots,\lambda_n)$ be a symmetric function of $n$ variables,
vanishing at $\lambda_j=\lambda_k$, $j,k=1,\dots,n$. Then,
\ba{GFpartlam}
&&{\dis\hspace{2mm}
\frac1{M^n}\sum_{\{\lambda\}=\{\lambda_{\alpha_+}\}\cup\{\lambda_{\alpha_-}\}
\atop{|\alpha_+|=n}}{\cal F}(\{\lambda_{\alpha_+}\})
=\frac1{n!M^n}\sum_{j_1=1}^N\cdots \sum_{j_n=1}^N
{\cal F}(\lambda_{j_1},\dots,\lambda_{j_n})}\non
&&{\dis\hspace{12mm}
\longrightarrow
\frac1{n!}
\int_C d\lambda_1\rho_{tot}(\lambda_1)
\cdots \int_C d\lambda_n\rho_{tot}(\lambda_n)
{\cal F}(\lambda_1,\dots,\lambda_n),\qquad \mbox{as}~ M\to\infty,}
\ea
when $C$ is the contour involved in the Lieb equation \eq{GFLiebeq}.

Thus, the sum with respect to partitions in the thermodynamic
limit transforms into $2n$ integrals, and we obtain the
expectation value \eq{GFebQm+1} of the operator $\exp(\beta Q_{1,m})$ in the
ground state.
\qed

We would like to mention that, as for the elementary blocks of correlation
functions obtained in \cite{KitMT00}, the
integrand in \eq{GFebQm+1hom} consists of two parts: thermodynamic
and algebraic. The thermodynamic part depends on the phase of the
model ($\Delta$ and $h$), but not on the considered set of local operators;
it includes the determinant of densities
and the integration contour $C$ for the variables $\{\lambda\}$.
On the contrary, the algebraic part, which includes the
remaining factors of the integrand, does not depend on the phase
of the model, but only on the particular set of operators the
correlation functions of which one wants to compute.

\bigskip

Nevertheless, there exists a principal difference between the
representation  \eq{GFebQm+1hom} and the multiple integrals for
the elementary blocks:

Each of the elementary blocks involved in the construction of
$\prod_{a=1}^{m}\left(A+e^\beta D\right)(\xi_a)$ contains
exactly $m$ integrals. However, some of these integrals, which
correspond to the action of $D(\xi_a)$, are taken over the
contours $C$ (see \eq{GFLiebeq}), while the other integration
contours (corresponding to $A$ type actions) are shifted.  The
reason of this difference is that, due to the fact that $d(\xi_a)=0$,
the operator $D(\xi_a)$ has no direct action, while $A(\xi_a)$ does have. The
shift of the contours allows us to get rid of the terms produced by
the direct action of $A(\xi_a)$ (see \cite{KitMT00} for more
details).

On the contrary, in the equation \eq{GFebQm+1hom}, the integrals
with respect to all $\{\lambda\}$ are taken over the same contour
$C$. Moreover, the number of these integrals is not fixed, but
varies from $0$ to $m$: the term $n=m$ in \eq{GFebQm+1hom}
corresponds to completely indirect actions of $A$ and $D$; the
terms with $n<m$ contain direct actions of $A$, and in particular
the term $n=0$ describes the direct action of the whole product
$A(\xi_1)\dots A(\xi_m)$.

From this observation, one can easily understand how to obtain the
representation \eq{GFebQm+1hom} from the formulas for the
elementary blocks. First of all, one should move back all the
shifted contours in each of the $2^m$ constituent elementary blocks
of the product $\prod_{a=1}^{m}\left(A+e^\beta D\right)(\xi_a)$.
Hereby one crosses the poles of the densities $\rho(\lambda,\xi)$,
which produces the terms where the number of integrals is less than
$m$. Then, one needs to gather the terms with the same number
of integrals and symmetrize all the obtained integrands with
respect to $\{\lambda\}$. The symmetrization produces the sum over
the partitions of inhomogeneities  $\{\xi\}$, which can be
effectively taken into account by the set of auxiliary
$z$-integrals. Generically, this way meets extremely serious
practical difficulties. However, for some simple particular cases,
it can be successfully applied. In Appendix \ref{EFP},  we
illustrate this method  by considering the limit
$\beta\to\infty$ in the equation \eq{GFebQm+1hom}. Then  we have
\be{GFlimexpbQ} \lim_{\beta\to\infty}e^{-\beta m} \exp(\beta
Q_{1,m})=\prod_{a=1}^{m}D(\xi_a) \prod_{b=1}^{m}(A+D)^{-1}(\xi_b).
\ee
Hence, we obtain the correlation function corresponding to the
emptiness formation probability which was considered in Section 4 of
\cite{KitMT00}. On the other hand, in the equation
\eq{GFebQm+1hom}, only the term with $n=m$ survives, and the
determinant $\det\tilde M_{jk}$ simplifies:
\be{GFlimM}
\lim_{\beta\to\infty}e^{-\beta m}
{\det}_{m}\tilde M_{jk}
=\prod_{a,b=1}^{m}\frac{\sinh(\lambda_b-z_a+\eta)}
{\sinh(z_a-\lambda_b+\eta)}
{\det}_{m}[t(\lambda_j,z_k)].
\ee
In Appendix \ref{EFP}, we obtain this result by symmetrization of
the corresponding multiple integral obtained in \cite{KitMT00}.
Finally we observe that, due to the fact that ${\det}_{n}\tilde M_{jk}
=\delta_{n0}$ at $\beta=0$, the expectation value of the identity
operator is equal to $1$. This obvious fact would be highly
nontrivial to prove from the multiple integral representations
for elementary blocks.


\section{Spin-spin correlation functions \label{ssc}}

In the previous section, we have obtained a multiple integral
representation of the correlation function
$\langle\sigma_1^z\sigma_{m+1}^z\rangle$ from the generating
functional $\langle\exp(\beta Q_{1,m})\rangle$. It has been computed
from the average value in the ground state of the product of $m$
commuting operators $(A + e^{\beta} D)(\xi_{\alpha})$ for $\alpha
= 1, \dots, m$. However, for general $n$-spin correlation functions,
we have to deal with ground state average values of
products of non-commuting operators. Indeed, a generic $k$-point
correlation function can be written as
\be{EEE} \langle\psi|\prod\limits_{j=1}^k
E^{\epsilon'_j,\epsilon_j}_{m_j}|\psi\rangle, \ee
where $m_1 < m_2 < \dots < m_k$ is an ordered set of $k$ sites on
the lattice. Using the solution of the quantum inverse scattering
problem, it can be reduced to the evaluation of the following
average value:
\be{TA+DTA+DT}
\langle\psi|T_{\epsilon_1,\epsilon'_1}(\xi_{m_1})\cdot\prod\limits_{\alpha
= m_1 + 1}^{m_2 - 1} (A + D)(\xi_{\alpha})\cdot
T_{\epsilon_2,\epsilon'_2}(\xi_{m_2})\cdot\prod\limits_{\alpha = m_2 +
1}^{m_3 - 1} (A + D)(\xi_{\alpha}) \dots
T_{\epsilon_k,\epsilon'_k}(\xi_{m_k})|\psi\rangle, \ee
To compute such a correlation function, the strategy is the
following. As usual, we act with the operators to the left.
The action of the shift operators (products of transfer matrices)
is given as in \eq{APcompl-act}. The action of the isolated
elements of the monodromy matrix
$T_{\epsilon_j,\epsilon'_j}(\xi_{m_j})$ for $j = 1, \dots, k$ is
computed as in the elementary blocks: in particular, in the
thermodynamic limit, we get rid of all direct type terms by
shifting the corresponding integration contours (see
\cite{KitMT00}). This helps the final expression to be as compact
as possible. We apply below this procedure to the case of the spin-spin
correlation functions.

\bigskip

Let us start with $\langle\sigma_1^z\sigma_{m+1}^z\rangle$.

It is possible to evaluate this quantity by computing directly
the expectation value
\be{Oexpszsz}
\langle\psi|\sigma_1^z\sigma_{m+1}^z|\psi\rangle=
\langle\psi|(A-D)(\xi_{1})\cdot
\prod\limits_{a=2}^{m}(A+D)(\xi_a)\cdot
(A-D)(\xi_{m+1})\cdot
\prod\limits_{b=1}^{m+1}(A+D)^{-1}(\xi_b)|\psi\rangle.
\ee
First, we has to to act with  $(A-D)(\xi_{1})$ on the eigenstate
$\langle\psi|$ using \eq{FCactA}, \eq{FCactD}; then, we can apply
\eq{APactxi} for the action of the product
$\prod_{a=2}^{m}(A+D)(\xi_a)$, and finally again use
\eq{FCactA}, \eq{FCactD} for the action of $(A-D)(\xi_{m+1})$.
This method does not meet some new principal obstacles, but the final answer
has a bit more complicated form. In fact, in this case, one obtains a sum
similar to \eq{GFebQm+1hom}, but in which the $n^{th}$ term
consists of four summands corresponding to the action of the operators $A$ and
$D$ in the points $\xi_1$ and $\xi_{m+1}$. Therefore the evaluation of
$\langle\sigma_1^z\sigma_{m+1}^z\rangle$ via the generating functional
seems to be the most preferable, at least from the stand point of its
compactness.

However, in particular cases, other representations for
the correlation function $\langle\sigma_1^z\sigma_{m+1}^z\rangle$
may present certain advantages, even if these representations have
slightly more complicated forms. Note that generically the correlation
function $\langle\sigma_1^z\sigma_{m+1}^z\rangle$ contains the term
$\langle\sigma^z\rangle^2$ which, due to the translation invariance,
does not depend on the distance $m$.
However, at zero magnetic field in the massless
regime, the magnetization $\langle\sigma^z\rangle$ is zero and the correlations
of the third components of spin should be a decreasing function of $m$.
On the other hand, each term of the sum \eq{GFebQm+1hom}, even after taking
the lattice and $\beta$-derivatives,
still contains a part which does not depend
on the distance $m$. It follows from \eq{GFlatder} that eventually these
constant terms give $1/2$. For the purposes of the forthcoming asymptotic
analysis, it would be desirable to obtain a representation of
$\langle\sigma_1^z\sigma_{m+1}^z\rangle$ without any constant
contribution at zero magnetic field. Such a representation, of course, is
provided by \eq{Oexpszsz}, but we can also consider a certain modification of
the generating functional $\langle\exp(\beta Q_{1,m})\rangle$, namely
$\langle\exp(\beta Q_{1,m})\sigma_{m+1}^z\rangle$. Indeed, it is easy to see
that
\be{O1-szsz}
\frac12\langle(1-\sigma_1^z)\sigma_{m+1}^z\rangle=\left.\frac{
\partial}{\partial\beta}
\langle\Bigl[\exp(\beta Q_{1,m})-\exp(\beta Q_{2,m})\Bigr]
\sigma_{m+1}^z\rangle
\right|_{\beta=0}~,
\ee
and, since the magnetization is zero at $h=0$ in the massless regime, we obtain
\be{Oszsz}
\langle\sigma_1^z\sigma_{m+1}^z\rangle=
\left.-2{\cal D}_m\frac{\partial}{\partial\beta}
\langle\exp(\beta Q_{1,m})\sigma_{m+1}^z\rangle
\right|_{\beta=0},\qquad h=0,\qquad |\Delta|<1.
\ee
\begin{prop}\label{sz-sz}
The ground state expectation value $\langle\exp(\beta
Q_{1,m})\sigma_{m+1}^z\rangle$ for the homogeneous case is given as
\ba{OnewGFmag}
&&{\dis\hspace{-7mm}
\langle\exp(\beta Q_{1,m})\sigma_{m+1}^z\rangle=
\sum_{n=0}^m\frac{-1}{(n!)^2}
\oint\limits_{\Gamma}
\prod_{j=1}^n\frac{dz_j}{2\pi i} \int_C d^{n}\lambda \cdot
\prod_{a=1}^n \left(\frac{\sinh(z_a+\frac\eta2)\sinh(\lambda_a-\frac\eta2)}
{\sinh(z_a-\frac\eta2)\sinh(\lambda_a+\frac\eta2)}\right)^{m}
}\non
&&{\hspace{14mm} \times
W_n(\{\lambda\}|\{z\})
{\det}_{n}\Bigl[\tilde M_{jk}(\{\lambda\}|\{z\})\Bigr]
\prod_{a=1}^n \frac{\sinh(\lambda_a-\frac\eta2)}
{\sinh(z_a-\frac\eta2)}}\num
&&{\dis\hspace{-7mm}
\times  \left(\int\limits_{\tilde C} d\lambda_{n+1}
\prod_{a=1}^n\frac{\sinh(\lambda_{n+1}-z_a-\eta)}
{\sinh(\lambda_{n+1}-\lambda_a-\eta)}
+\int\limits_C d\lambda_{n+1}
\prod_{a=1}^n\frac{\sinh(\lambda_{n+1}-z_a+\eta)}
{\sinh(\lambda_{n+1}-\lambda_a+\eta)}\right)
{\det}_{n+1}\Bigl[\rho(\lambda_j,z_k)\Bigr].}\nonumber
\ea
Here one should set $z_{n+1}=\eta/2$ in the last column of the determinant
of the densities.  $C$ and $\Gamma$ are as in Proposition \ref{Q},
while the shifted contour $\tilde C$ for $\lambda_{n+1}$ is
such that $\tilde C\cup(-C)$ surrounds the points $\{z\}$, in which
the functions $\rho(\lambda_{n+1},z_k)$ have simple poles, but it does not
contain other singularities.
For zero magnetic field one has
\ba{OnewGF}
&&{\dis\hspace{-2mm}
\langle\exp(\beta Q_{1,m})\sigma_{m+1}^z\rangle=
\sum_{n=0}^m\frac1{(n!)^2}
\oint\limits_{\Gamma}
\prod_{j=1}^n\frac{dz_j}{2\pi i} \int_{C} d^{n+1}\lambda \cdot
\prod_{a=1}^n \left(\frac{\sinh(z_a+\frac\eta2)\sinh(\lambda_a-\frac\eta2)}
{\sinh(z_a-\frac\eta2)\sinh(\lambda_a+\frac\eta2)}\right)^{m}
}\non
&&{\hspace{-2mm} \times
\prod_{a=1}^n \left(\frac{\sinh(\lambda_a-\frac\eta2)}
{\sinh(z_a-\frac\eta2)}\right)
\left(\prod_{a=1}^n \frac{\sinh(\lambda_{n+1}-z_a)}
{\sinh(\lambda_{n+1}-\lambda_a)} - \prod_{a=1}^n
\frac{\sinh(\lambda_{n+1}-z_a+\eta)}
{\sinh(\lambda_{n+1}-\lambda_a+\eta)}\right)\cdot
W_n(\{\lambda\}|\{z\})}\non
&&{\dis\hspace{8mm}
\times
{\det}_{n}\Bigl[\tilde M_{jk}(\{\lambda\}|\{z\})\Bigr]
{\det}_{n+1}\Bigl[\rho(\lambda_j,z_1),\dots,
\rho(\lambda_j,z_n),
\rho(\lambda_j,{\textstyle\frac{\eta}{2}})\Bigr].}
\ea
\end{prop}
{\sl Proof.}~~We ommit parts of the proof which coincide with
parts of the proof of Proposition \ref{Q}. Instead, we focus our
attention on certain peculiarities.

The normalized expectation value of $\exp(\beta
Q_{1,m})\sigma_{m+1}^z$ on the finite lattice is given by
\be{OexpbQszev}
\frac{\langle\psi|\exp(\beta Q_{1,m})\sigma_{m+1}^z|\psi\rangle}
{\langle\psi|\psi\rangle}=
\frac{\langle\psi|
\prod\limits_{a=1}^{m}(A+e^\beta D)(\xi_a)\cdot
(A-D)(\xi_{m+1})\cdot
\prod\limits_{b=1}^{m+1}(A+D)^{-1}(\xi_b)|\psi\rangle}
{\langle\psi|\psi\rangle}.
\ee
After the calculation of the actions of the product
$\prod\limits_{b=1}^{m+1}(A+D)^{-1}(\xi_b)$  on the state $|\psi\rangle$
and of $\prod_{a=1}^{m}(A+e^\beta D)(\xi_a)$ on
the state $\langle\psi|$, one has to act with
$(A-D)(\xi_{m+1})$ on the resulting states
$\langle 0 |\prod_{a\in\gamma_+} C(\xi_a)
\prod_{b\in\alpha_-}C(\lambda_b)$.

The action of $D(\xi_{m+1})$ gives
\ba{OactD}
&&{\dis\hspace{-14mm}
\langle 0 |\prod_{a\in\gamma_+} C(\xi_a)\prod_{b\in\alpha_-}C(\lambda_b)
\cdot D(\xi_{m+1})=\sum_{\ell\in\alpha_-}
\frac{\sinh\eta}{\sinh(\lambda_\ell-\xi_{m+1})}
\prod_{a\in\gamma_+}f(\lambda_\ell,\xi_a)
\prod_{a\in\alpha_-\atop{a\ne\ell}}f(\lambda_a,\lambda_\ell)}\non
&&{\dis\hspace{35mm}
\times
\prod_{a\in\alpha_+}\frac{f(\lambda_a,\lambda_\ell)}
{f(\lambda_\ell,\lambda_a)}\cdot
\langle 0 | C(\xi_{m+1})\prod_{a\in\gamma_+} C(\xi_a)
\prod_{b\in\alpha_-\atop{b\ne\ell}}C(\lambda_b).}
\ea
Here we have used that $\{\lambda\}$ satisfy the system of Bethe
equations.  Then, one has to compute the normalized scalar products of
the obtained states with $|\psi\rangle$ and proceed to the thermodynamic
limit. We obtain:
\ba{OactDtd}
&&{\dis\hspace{-5mm}
\langle\exp(\beta Q_{1,m})D(\xi_{m+1})\rangle=
\sum_{n=0}^m\frac1{(n!)^2}
\oint\limits_{\Gamma}
\prod_{j=1}^n\frac{dz_j}{2\pi i} \int_C d^{n}\lambda \cdot
\prod_{b=1}^n\prod_{a=1}^{m} \frac{f(z_b,\xi_a)}
{ f(\lambda_b,\xi_a)}
}\non
&&{\hspace{14mm} \times
W_n(\{\lambda\}|\{z\})
{\det}_{n}\Bigl[\tilde M_{jk}(\{\lambda\}|\{z\})\Bigr]
\prod_{a=1}^n \frac{\sinh(\lambda_a-\xi_{m+1})}
{\sinh(z_a-\xi_{m+1})}}\non
&&{\dis\hspace{-5mm}
\times  \int_C d\lambda_{n+1}\cdot
\prod_{a=1}^n\frac{\sinh(\lambda_{n+1}-z_a+\eta)}
{\sinh(\lambda_{n+1}-\lambda_a+\eta)}\cdot
{\det}_{n+1}\Bigl[\rho(\lambda_j,z_1),\dots,
\rho(\lambda_j,z_n),
\rho(\lambda_j,\xi_{m+1})\Bigr].}
\ea
Observe that, in comparison with \eq{GFebQm+1hom}, the integrand in
\eq{OactDtd} contains additional factors. Moreover, we see
that the sum over $\ell$ in \eq{OactD} produces one more integral
with respect to $\lambda_{n+1}$ in the thermodynamic limit.

The action of the operator $A(\xi_{m+1})$ is more complicated. First,
it contains terms similar to \eq{OactD}:
$$
\sum_{\ell\in\alpha_-}
\frac{\sinh\eta}{\sinh(\xi_{m+1}-\lambda_\ell)}
\prod_{a\in\gamma_+}f(\xi_a,\lambda_\ell)
\prod_{a\in\alpha_-\atop{a\ne\ell}}f(\lambda_a,\lambda_\ell)\cdot
\langle 0 | C(\xi_{m+1})\prod_{a\in\gamma_+} C(\xi_a)
\prod_{b\in\alpha_-\atop{b\ne\ell}}C(\lambda_b).
$$
In the thermodynamic limit, the contribution of these terms turns into
integrals of the type \eq{OactDtd}, where in the last line one
should make the replacement
$$
\prod_{a=1}^n\frac{\sinh(\lambda_{n+1}-z_a+\eta)}
{\sinh(\lambda_{n+1}-\lambda_a+\eta)}\longrightarrow-
\prod_{a=1}^n\frac{\sinh(\lambda_{n+1}-z_a-\eta)}
{\sinh(\lambda_{n+1}-\lambda_a-\eta)}.
$$
However, the action of the operator $A(\xi_{m+1})$ contains also
the direct term
$$
\prod_{a\in\gamma_+}f(\xi_a,\xi_{m+1})
\prod_{a\in\alpha_-}f(\lambda_a,\xi_{m+1})\cdot
\langle 0 |\prod_{a\in\gamma_+} C(\xi_a)
\prod_{b\in\alpha_-}C(\lambda_b),
$$
and terms where the argument $\xi_{m+1}$ exchanges with one
of $\{\xi_{\gamma_+}\}$:
$$
\sum_{\ell\in\gamma_+}
\frac{\sinh\eta}{\sinh(\xi_{m+1}-\xi_\ell)}
\prod_{a\in\gamma_+\atop{a\ne\ell}}f(\xi_a,\xi_\ell)
\prod_{a\in\alpha_-}f(\lambda_a,\xi_\ell)\cdot
\langle 0 | C(\xi_{m+1})\prod_{a\in\gamma_+\atop{a\ne\ell}} C(\xi_a)
\prod_{b\in\alpha_-}C(\lambda_b).
$$
These terms give also non-vanishing contributions to the
expectation value $\langle\exp(\beta
Q_{1,m})\sigma_{m+1}^z\rangle$. However, in the thermodynamic
limit, one can withdraw them by shifting the integration contour
for $\lambda_{n+1}$ in complete analogy with the method used in
\cite{KitMT00} for the elementary blocks. This gives us
\eq{OnewGFmag}. This representation is valid in arbitrary regime
of the model. For zero magnetic field, the original integration
contour $C$  is the real axis for $|\Delta|<1$ and the interval
$[i\pi/2,-i\pi/2]$ for $\Delta>1$. In both cases we can choose
$\tilde C=C+\eta$ (recall that in the massless regime
$\eta=-i\zeta$, $\zeta>0$). Changing then $\lambda_{n+1}$ with
$\lambda_{n+1}+\eta$, we arrive at \eq{OnewGF}. \qed

The representation \eq{OnewGF}, as it was expected, has a slightly more
complicated form than \eq{GFebQm+1hom}. However, after the
lattice and $\beta$-derivations, the sum \eq{OnewGF} does not
contain any constant contribution for $|\Delta|<1$. This fact may play
an important role for the asymptotic analysis of the correlation
function $\langle\sigma_1^z\sigma_{m+1}^z\rangle$.

\bigskip

The remaining two-point functions are
$\langle\sigma_1^-\sigma_{m+1}^+\rangle$ and
$\langle\sigma_1^+\sigma_{m+1}^-\rangle$. On the finite lattice
these two quantities are given by
\be{O-+ev}
\frac{\langle\psi|\sigma_1^-\sigma_{m+1}^+|\psi\rangle}
{\langle\psi|\psi\rangle}
=\frac{\langle\psi|B(\xi_1)\cdot
\prod\limits_{a=2}^{m}(A+D)(\xi_a)\cdot C(\xi_{m+1})\cdot
\prod\limits_{b=1}^{m+1}(A+D)^{-1}(\xi_b)|\psi\rangle}
{\langle\psi|\psi\rangle},
\ee
\be{O+-ev}
\frac{\langle\psi|\sigma_1^+\sigma_{m+1}^-|\psi\rangle}
{\langle\psi|\psi\rangle}
=\frac{\langle\psi|C(\xi_1)\cdot
\prod\limits_{a=2}^{m}(A+D)(\xi_a)\cdot B(\xi_{m+1})\cdot
\prod\limits_{b=1}^{m+1}(A+D)^{-1}(\xi_b)|\psi\rangle}
{\langle\psi|\psi\rangle}
\ee
It is clear that our method can be applied without significant
changes for the calculation of \eq{O-+ev}, \eq{O+-ev} as well.
Therefore we present here only the final result. Moreover, for
simplicity, we first consider the case of zero magnetic field (for
both massive and massless phases), when the correlation functions
$\langle\sigma_1^-\sigma_{m+1}^+\rangle$ and
$\langle\sigma_1^+\sigma_{m+1}^-\rangle$ coincide.
\begin{prop}\label{sp-sm}
The ground state expectation value $\langle\sigma^+_1
\sigma^-_{m+1}\rangle$ for the homogeneous case
at zero magnetic field can be expressed as
\ba{Os+s-}
&&{\dis\hspace{-7mm}
\langle \sigma^+_1\sigma^-_{m+1}\rangle=\sum_{n=0}^{m-1}
\frac1{n!(n+1)!}\oint\limits_{\Gamma}
 \prod_{j=1}^{n+1}\frac{dz_j}{2\pi i}
\int_C d^{n+2}\lambda
\prod_{a=1}^{n+1} \left(\frac{\sinh(z_a+\frac\eta2)}{\sinh(z_a-\frac\eta2)}
\right)^m \hspace{2mm}
\prod_{a=1}^{n}\left(\frac{\sinh(\lambda_a-\frac\eta2)}
{\sinh(\lambda_a+\frac\eta2)}\right)^{m}
}\non
&&{\dis\hspace{7mm}
\times \frac1{\sinh(\lambda_{n+1}-\lambda_{n+2})}\cdot
\left(\frac{\prod\limits_{a=1}^{n+1}
\sinh(\lambda_{n+1}-z_a+\eta)\sinh(\lambda_{n+2}-z_a)}
{\prod\limits_{a=1}^n\sinh(\lambda_{n+1}-\lambda_a+\eta)
\sinh(\lambda_{n+2}-\lambda_a)}
\right)\cdot \hat W_n(\{\lambda\},\{z\})}\non
&&{\dis\hspace{7mm}
\times {\det}_{n+1} \hat M_{jk}\cdot
{\det}_{n+2}\left[
\rho(\lambda_j,z_1),\dots,\rho(\lambda_j,z_{n+1}),
\rho(\lambda_j,{\textstyle\frac{\eta}{2}})
\right],}
\ea
where the contours $C$ and $\Gamma$ are defined as in Proposition \ref{Q}.
Here the analog of the function $W_n(\{\lambda\},\{z\})$
is
\be{OhWn}
\hat W_n(\{\lambda\},\{z\})=
\frac{\prod\limits_{a=1}^n\prod\limits_{b=1}^{n+1}
\sinh(\lambda_a-z_b+\eta)\sinh(z_b-\lambda_a+\eta)}
{\prod\limits_{a=1}^n\prod\limits_{b=1}^n
\sinh(\lambda_a-\lambda_b+\eta)
\prod\limits_{a=1}^{n+1}\prod\limits_{b=1}^{n+1}
\sinh(z_a-z_b+\eta)},
\ee
and the $(n+1)\times(n+1)$ matrix $\hat M$ has
the entries
\be{OhM}
\hat M_{jk}=t(z_k,\lambda_j)-t(\lambda_j,z_k)
\prod_{a=1}^n\frac{\sinh(\lambda_a-\lambda_j+\eta)}
{\sinh(\lambda_j-\lambda_a+\eta)}
\prod_{b=1}^{n+1}\frac{\sinh(\lambda_j-z_b+\eta)}
{\sinh(z_b-\lambda_j+\eta)},
\qquad j\le n,
\ee
and $\hat M_{n+1,k}=t(z_k,{\textstyle \frac\eta2})$ for $j=n+1$.
\end{prop}
{\sl Proof.}~~ The only difference between \eq{O+-ev} and the
expectation values considered above is that now we deal with the
operators $C$ and $B$, and thus use \eq{FCactB} for
the action of $B$ (recall that the action of $C$ is free). The
rest of the computations is mostly the same. In analogy with
\eq{OnewGF}, the integrals with respect to $\lambda_{n+1}$ and
$\lambda_{n+2}$ describe the action of the operator $B(\xi_{m+1})$
in \eq{O+-ev}. The direct action of $B(\xi_{m+1})$ is taken into
account by the shift of the integration contour for
$\lambda_{n+2}$.
\qed

It is clear that this result can be easily generalized for the
case of non-zero magnetic field in complete analogy with
\eq{OnewGFmag}. In particular, for the correlation function
$\langle \sigma^+_1\sigma^-_{m+1}\rangle$, one should replace
the variable $\lambda_{n+2}$ in
\eq{Os+s-} with $\lambda_{n+2}+\eta$, and choose
for this variable the shifted integration contour $\tilde C$.


\section*{Conclusion}

The main result of this paper is a new multiple integral
representation for the spin-spin correlation functions at lattice
distance $m$ of the $XXZ$ Heisenberg chain in a magnetic field. In
particular, it gives generically  an effective re-summation of the
corresponding $2^m$ elementary blocks as the sum of only $m$
terms, each containing the distance as the power $m$ of some
simple function. Hence, our method opens the possibility of the
asymptotic analysis of the spin-spin correlation functions at
large distance. It will be shown in a separate publication that it
also leads in a direct way to the known answers at the free
fermion point $\Delta = 0$.

It should also be noted that the compact formula for the multiple
action of the transfer matrix operator, for any values of the
spectral parameter and on arbitrary quantum state, is central in
our result. It can be used to compute multi-spins correlation
functions. It contains in particular the possibility to act on any
quantum state with generic conserved quantities responsible for
the quantum integrability of the $XXZ$ Heisenberg chain, for
example with the Hamiltonian itself. Therefore, this formula is
also the key to the dynamical correlation functions. Note finally
that this result depends only on the general structure of the
$R$-matrix, and thus can be generalized to other models admitting
quantum inverse scattering problem solution \cite{MaiT00}, like
the integrable Heisenberg higher spin chains \cite{Kit01}.

\section*{Acknowledgments}

N. K. would like to thank the University of York, the SPhT in
Saclay and JSPS  for financial support. N. S. is supported by the
grants INTAS-99-1782, RFBR-99-01-00151, Leading Scientific Schools
00-15-96046, the Program Nonlinear Dynamics and Solitons and
CNRS. J.M. M. is supported by CNRS. V. T is supported by DOE grant
DE-FG02-96ER40959 and by CNRS.  N. K, N. S. and V. T. would like
to thank the Theoretical Physics group of the Laboratory of
Physics at ENS Lyon for hospitality, which makes this collaboration
possible.
\appendix
\section{The highest coefficient \label{HC}}
Let
\be{HCSn}
S_n(x_1,\dots,x_n|\mu_1,\dots,\mu_n|\mu_{n+1},\dots,\mu_N)=
\frac{\prod\limits_{b=1}^n\prod\limits_{a=1}^n
\sinh(x_a-\mu_b+\eta)}
{\prod\limits_{a>b}^n
\sinh(\mu_a-\mu_b)
\sinh(x_b-x_a)}\cdot
{\det}_n M_{jk},
\ee
where the $n\times n$ matrix $M_{jk}$ is
\be{HCMjk}
M_{jk}=a(\mu_j)t(x_k,\mu_j)
\prod_{a=n+1}^N
f(\mu_a,\mu_j)
-e^\beta d(\mu_j)t(\mu_j,x_k)
\prod_{a=n+1}^Nf(\mu_j,\mu_a)
\prod_{b=1}^n\frac{\sinh(\mu_j-x_b+\eta)}
{\sinh(\mu_j-x_b-\eta)}.
\ee
First of all we prove an auxiliary lemma, establishing the recursion property
of the function $S_n$.
\begin{lemma}\label{recur}
\ba{HCprop}
&&{\dis\hspace{-7mm}
S_n(x_1,\dots,x_n|\mu_1,\dots,\mu_n|\mu_{n+1},\dots,\mu_N)}\non
&&{\dis\hspace{11mm}
=\sum_{l=1}^nS_{n-1}(x_1,\dots,x_{n-1}|
\mu_1,\dots,\check\mu_l,\dots,\mu_n|\mu_{n+1},\dots,\mu_N,\mu_l)}\num
&&{\dis\hspace{-8mm}
\times\left(
a(\mu_l)g(x_{n},\mu_l)\prod_{a=1}^{n-1}f(x_a,\mu_l)
\prod_{a=n+1}^Nf(\mu_a,\mu_l)+
e^\beta d(\mu_l)g(\mu_l,x_{n})\prod_{a=1}^{n-1}f(\mu_l,x_a)
\prod_{a=n+1}^Nf(\mu_l,\mu_a)
\right),}\nonumber
\ea
where the symbol $\check\mu_l$ means that the corresponding parameter
is ommited in the set $\mu_1,\dots,\mu_n$.
\end{lemma}
{\sl Proof.}~~Consider an auxiliary contour integral:
\be{HCauxint}
I=\frac1{2\pi i}\int\frac{d\omega}{\sinh(x_n-\omega)}
S_n(x_1,\dots,x_{n-1},\omega|\mu_1,\dots,\mu_n|\mu_{n+1},\dots,\mu_N).
\ee
The integral is taken with respect to the boundaries of a
horizontal strip of the  width $i\pi$. For instance, one can take
for the lower boundary $\Im(\omega)=\omega_0$, and for the upper
boundary $\Im(\omega)=\omega_0+i\pi$. Hereby $\omega_0$ is
an arbitrary real number satisfying the conditions $\omega_0\ne
\Im(x_n+i\pi k)$ and $\omega_0\ne \Im(\mu_j+i\pi k)$, where
$j=1,\dots,n$, $k\in\mathbb{Z}$. Obviously, the integrand
decreases as $\exp(-2|\omega|)$ at $\omega\to \pm\infty$.
Moreover, the integrand is a periodic function of $\omega$ with
the period $i\pi$. Thus, $I=0$ and, hence, the sum of the residues
inside the contour vanishes. The pole at $\sinh(\omega-x_n)=0$
gives us the term in the l.h.s. of \eq{HCprop}. On the other hand,
the only singularities of the function $S_n$ are simple poles at
$\sinh(\omega-\mu_j)=0$. The residues in these poles give us the
r.h.s. of \eq{HCprop}. Thus, the lemma is proved.
\qed

\begin{prop}
The highest coefficient of the completely indirect action is equal to
the function $S_n$ \eq{HCSn} at $\mu_1,\dots,\mu_n=\{\lambda_{\alpha_+}\}$
and $\mu_{n+1},\dots,\mu_N=\{\lambda_{\alpha_-}\}$.
\end{prop}
{\sl Proof.}~~
One can use the induction with respect to $n$.
For $n=1$, the equations \eq{APSm}, \eq{APMjk} give us exactly \eq{APactsing}.
Let the highest coefficient have the form \eq{APSm}, \eq{APMjk} for $n-1$. Then
we have
\ba{HCind1}
&&{\dis\hspace{-11mm}
\langle\psi|
\left.\prod_{a=1}^n\left(A+e^\beta D\right)(x_a)\right|_{(c.-ind.)}
=\sum_{\{\lambda\}=\{\lambda_{\alpha_+}\}\cup\{\lambda_{\alpha_-}\}
\atop{|\alpha_+|=n-1}}
S_{n-1}(x_1,\dots,x_{n-1}|\{\lambda_{\alpha_+}\}|\{\lambda_{\alpha_-}\})
}\non
&&{\dis\hspace{14mm}
\times
\langle 0| \prod_{a=1}^{n-1}C(x_a)
\left.\prod_{b\in\alpha_-}C(\lambda_b)\cdot
\left(A+e^\beta D\right)(x_n)\right|_{(c.-ind.)}.}
\ea
Now, acting with the last operator, we need to exchange $x_n$ with one of
the $\lambda\in\{\lambda_{\alpha_-}\}$. This gives us
\ba{HCind2}
&&{\dis\hspace{-5mm}
\langle\psi|
\left.\prod_{a=1}^n\left(A+e^\beta D\right)(x_a)\right|_{(c.-ind.)}
=\sum_{\{\lambda\}=\{\lambda_{\alpha_+}\}\cup\{\lambda_{\alpha_-}\}
\atop{|\alpha_+|=n-1}}\sum_{l\in\alpha_-}
S_{n-1}(x_1,\dots,x_{n-1}|\{\lambda_{\alpha_+}\}|\{\lambda_{\alpha_-}\})
}\non
&&{\dis\hspace{-5mm}
\times\left(
a(\lambda_l)g(x_{n},\lambda_l)\prod_{a=1}^{n-1}f(x_a,\lambda_l)
\prod_{a\in\alpha_-\atop{a\ne l}}f(\lambda_a,\lambda_l)+
e^\beta d(\lambda_l)g(\lambda_l,x_{n})\prod_{a=1}^{n-1}f(\lambda_l,x_a)
\prod_{a\in\alpha_-\atop{a\ne l}}f(\lambda_l,\lambda_a)
\right)}\non
&&{\dis\hspace{54mm}
\times\langle 0| \prod_{a=1}^{n}C(x_a)
\prod_{b\in\alpha_-\atop{b\ne l}}C(\lambda_b).}
\ea
For each fixed partition, one can define the new sets
\be{HCnewpart}
\begin{array}{l}
\{\lambda_{\alpha'_+}\}=\{\lambda_{\alpha_+}\}\cup \lambda_l,\\
\{\lambda_{\alpha'_-}\}=\{\lambda_{\alpha_-}\}\setminus \lambda_l.
\end{array}
\ee
Then, \eq{HCind2} takes the form
\ba{HCind3}
&&{\dis\hspace{5mm}
\langle\psi|
\left.\prod_{a=1}^n\left(A+e^\beta D\right)(x_a)\right|_{(c.-ind.)}}\non
&&{\dis\hspace{15mm}
=\sum_{\{\lambda\}=\{\lambda_{\alpha'_+}\}\cup\{\lambda_{\alpha'_-}\}
\atop{|\alpha'_+|=n}}\hspace{4mm}\sum_{l\in\alpha'_+}
S_{n-1}(x_1,\dots,x_{n-1}|\{\lambda_{\alpha'_+}\}\setminus\lambda_l|
\{\lambda_{\alpha'_-}\}\cup\lambda_l)}\non
&&{\dis\hspace{-5mm}
\times\left(
a(\lambda_l)g(x_{n},\lambda_l)\prod_{a=1}^{n-1}f(x_a,\lambda_l)
\prod_{a\in\alpha'_-}f(\lambda_a,\lambda_l)+
e^\beta d(\lambda_l)g(\lambda_l,x_{n})\prod_{a=1}^{n-1}f(\lambda_l,x_a)
\prod_{a\in\alpha'_-}f(\lambda_l,\lambda_a)
\right)}\non
&&{\dis\hspace{54mm}
\times\langle 0| \prod_{a=1}^{n}C(x_a)
\prod_{b\in\alpha'_-}C(\lambda_b)}.
\ea
Identifying in  \eq{HCind3} and \eq{HCprop} $\lambda_l=\mu_l$,
$\{\lambda_{\alpha'_+}\}=\mu_1,\dots,\mu_n$ and
$\{\lambda_{\alpha'_-}\}= \mu_{n+1},\dots,\mu_N$, we come to the
conclusion that the coefficient at the state $\langle 0|
\prod_{a=1}^{n}C(x_a)\prod_{b\in\alpha'_-}C(\lambda_b)$ is exactly
$S_{n}(x_1,\dots,x_{n}|\{\lambda_{\alpha'_+}\}|
\{\lambda_{\alpha'_-}\})$, what ends the proof.
\qed


\section{The properties of the function $\tilde S_n$
\label{Sn}}

\begin{prop}
The matrix $\tilde M_{jk}(\xi_1,\dots,\xi_n|\lambda_1,\dots,\lambda_n)$
\eq{APtiMjk} at $\beta=0$ possesses the eigenvector
\be{PSNev}
\theta_j=\prod_{a=1}^n\sinh(\xi_j-\lambda_a)
\left(\prod_{a=1\atop{a\ne j}}^n\sinh(\xi_j-\xi_a)\right)^{-1}
\ee
with zero eigenvalue.
\end{prop}
{\sl Proof.}~~ The action of $\tilde
M_{jk}(\xi_1,\dots,\xi_n|\lambda_1,\dots,\lambda_n)$ at $\beta=0$ on the vector
\eq{PSNev} can be written in the form
\be{PSNact}
\sum_{k=1}^n\tilde M_{jk}\theta_k=
G_j^{(+)}+
G_j^{(-)}
\prod_{a=1}^n\frac{\sinh(\lambda_a-\lambda_j+\eta)\sinh(\lambda_j-\xi_a+\eta)}
{\sinh(\lambda_j-\lambda_a+\eta)\sinh(\xi_a-\lambda_j+\eta)},
\ee
where
\be{PSNG+-}
G_j^{(\pm)}=\sum_{k=1}^n\hspace{3mm}
\frac{\sinh\eta}{\sinh(\xi_k-\lambda_j)\sinh(\xi_k-\lambda_j\pm\eta)}
\cdot\frac{\prod\limits_{a=1}^n\sinh(\xi_k-\lambda_a)}
{\prod\limits_{a=1\atop{a\ne k}}^n\sinh(\xi_k-\xi_a)}.
\ee
To find $G_j^{(\pm)}$ we consider a contour integral similar to the
integral in Lemma \ref{recur}:
\be{PSNG+-int}
I_j^{(\pm)}=\frac1{2\pi i}\int\,
\frac{\sinh\eta }{\sinh(\omega-\lambda_j)
\sinh(\omega-\lambda_j\pm\eta)}
\cdot\prod\limits_{a=1}^n\frac{\sinh(\omega-\lambda_a)}
{\sinh(\omega-\xi_a)}\,d\omega.
\ee
Just like in the Lemma \ref{recur}, the  integral is taken with respect to the
boundaries of a horizontal strip of the  width $i\pi$.
Due to the periodicity of the integrand and its vanishing at
$\omega\to\pm\infty$, we conclude that $I_j^{(\pm)}=0$, and thus that
the sum of the residues within the contour vanishes. The sum of the residues
at $\sinh(\omega-\xi_k)=0$ gives $G_j^{(\pm)}$.  In addition, we have one more
pole at $\sinh(\omega-\lambda_j\pm\eta)=0$. Combining all together we find
\be{PSNG+-res}
G_j^{(\pm)}=\pm\prod_{a=1}^n\frac{\sinh(\lambda_a-\lambda_j\pm\eta)}
{\sinh(\xi_a-\lambda_j\pm\eta)}.
\ee
Substituting this into \eq{PSNact} we obtain $\sum_{k=1}^n\tilde
M_{jk}\theta_k=0$, and the Proposition is proved.
\qed

Thus, at $\beta=0$, the determinant of the matrix $\tilde M_{jk}$
vanishes for $n\ge1$ and, hence, $\tilde S_n=\delta_{n0}$.


\section{Symmetrization of an elementary block \label{EFP}}

The multiple integral obtained in \cite{KitMT00} for the emptiness formation
probability $\tau(m)$ on the inhomogeneous lattice has the form
\be{Stau-m-rep}
\tau(m)=\int_C
\frac{I(\{\lambda\},\{\xi\})}
{\prod\limits_{a<b}^m
\sinh(\xi_a-\xi_b)}
{\det}_m[\rho(\lambda_j,\xi_k)]\,d^m\lambda ,
\ee
where
\be{SI}
I(\{\lambda\},\{\xi\})=
\frac{\prod\limits_{j=1}^{m}\left\{\prod\limits_{k=1}^{j-1}
\sinh(\lambda_{j}-\xi_k+\eta)
\prod\limits_{k=j+1}^{m}
\sinh(\lambda_{j}-\xi_k)\right\}}
{\prod\limits_{a>b}^m
\sinh(\lambda_{a}-\lambda_{b}+\eta)}.
\ee
Clearly, due to the factor ${\det}_m[\rho(\lambda_j,\xi_k)]$,
the symmetrization of the integrand with respect to all $\{\lambda\}$ is
equivalent to the alternating sum of $I(\{\lambda\},\{\xi\})$ with
respect to the permutations  $\sigma:\lambda_1,\dots,\lambda_m\to
\lambda_{\sigma(1)},\dots,\lambda_{\sigma(m)}$.
\begin{prop}\label{Symmetr}
\be{Slemma}
\sum_{\sigma}(-1)^{p(\sigma)}
I(\{\lambda_{\sigma}\},\{\xi\})=
Z_m(\{\lambda\},\{\xi\}),
\ee
where
\be{S2Zm}
Z_m(\{\lambda\},\{\xi\})=
\prod\limits_{a=1}^{m}\prod\limits_{b=1}^{m}
\frac{\sinh(\lambda_a-\xi_b)\sinh(\lambda_a-\xi_b+\eta)}
{\sinh(\lambda_a-\lambda_b+\eta)}
\cdot\frac{{\det}_m[t(\lambda_j,\xi_k)]}{
\prod\limits_{a>b}^m\sinh(\xi_a-\xi_b)}.
\ee
\end{prop}
{\sl Proof.}~~It is convenient to introduce new variables
$x_j=e^{2\lambda_j},~  y_j=e^{2\xi_j},~q=e^{\eta}$. Then \eq{Slemma} takes
the form
\be{Sidentrat}
\sum_{\sigma}(-1)^{p(\sigma)}
\frac{\prod\limits_{j=1}^{m}\left\{\prod\limits_{k=1}^{j-1}
(q^{-1}x_{\sigma(j)}-qy_k)
\prod\limits_{k=j+1}^{m}
(x_{\sigma(j)}-y_k)\right\}}
{\prod\limits_{m\ge a>b\ge1}
(q^{-1}x_{\sigma(a)}-qx_{\sigma(b)})}
=\tilde Z_m(\{x\},\{y\}),
\ee
and
\ba{SZmrat}
&&{\dis\hspace{-1cm}
\tilde Z_m(\{x\},\{y\})=\left(\prod_{a=1}^{m}x_a\right)
\left(\prod\limits_{m\ge a>b\ge1}(y_a-y_b)\right)^{-1}
\prod\limits_{a=1}^{m}\prod\limits_{b=1}^{m}\left(
\frac{(x_a-y_b)(q^{-1}x_a-qy_b)}
{(q^{-1}x_a-qx_b)}\right)}\nona{35}
&&\hspace{5cm}{\dis\times
{\det}_m\left[\frac{q^{-1}-q}
{(x_j-y_k)(q^{-1}x_j-y_k)}\right].}
\ea
Observe that \eq{Sidentrat} holds for $m=1$. Suppose it is valid
for $m-1$.
Let us consider the properties of the both sides of \eq{Sidentrat} as
functions of  $y_m$. Obviously, these functions are polynomials of $y_m$ of
$m-1$ degree (the poles in  $y_m=y_a$ in the r.h.s. disappear due to the
zeros of the determinant in the same points).
The coefficients of these polynomials are antisymmetric
functions of parameters $\{x\}$.
Thus, in order to prove \eq{Sidentrat}, it is enough
to compare the values of both sides of this equality in
$y_m=x_a$, $a=1,\dots,m$. Moreover, due to the antisymmetry of the
coefficients with respect to $\{x\}$,
it is sufficient to consider the case $y_m=x_m$.
For $y_m=x_m$ the determinant in \eq{SZmrat} reduces to the product of
the last diagonal element by the corresponding minor. Extracting all the
dependency  on $x_m$ and $y_m$, we obtain for $\tilde Z_m$:
\be{SZmres}
\left.\vphantom{\int}\tilde Z_m(\{x\},\{y\})
\right|_{y_m=x_m}
=\prod_{a=1}^{m-1}
\frac{(x_a-y_m)(q^{-1}x_m-qy_a)}{(q^{-1}x_m-qx_a)}
\tilde Z_{m-1}(\{x\ne x_m\},\{y\ne y_m\}).
\ee
Consider now the l.h.s. of \eq{Sidentrat}. Each term of this
sum contains the product $\prod_{j=1}^{m-1}(x_{\sigma(j)}-y_m)$.
For $y_m=x_m$, this product does not vanish if and only if
$x_{\sigma(m)}=x_m$. Hence, in this case, we need to sum up only
with respect to the permutations of the $m-1$ variables $x_1,\dots,x_{m-1}$,
while $x_m$ remains fixed. Denoting these permutation as
$\sigma'$, and extracting again the
dependency on $x_m$ and $y_m$, we obtain for the l.h.s. \eq{Sidentrat}:
\be{Sinter}
\prod_{a=1}^{m-1}
\frac{(x_a-y_m)(q^{-1}x_m-qy_a)}{(q^{-1}x_m-qx_a)}
\sum_{\sigma'}(-1)^{p(\sigma')}
\frac{\prod\limits_{j=1}^{m-1}\left\{\prod\limits_{k=1}^{j-1}
(q^{-1}x_{\sigma'(j)}-qy_k)
\prod\limits_{k=j+1}^{m-1}
(x_{\sigma'(j)}-y_k)\right\}}
{\prod\limits_{m-1\ge a>b\ge1}
(q^{-1}x_{\sigma'(a)}-qx_{\sigma'(b)})}.
\ee
Due to the  assumption of the induction, the sum with respect to
the permutations of the variables $x_1,\dots,x_{m-1}$
gives $\tilde Z_{m-1}(\{x\ne
x_m\},\{y\ne y_m\})$. Then, comparison of \eq{Sinter} and
\eq{SZmres} completes the proof.
\qed

Thus, after symmetrization of the integrand, the multiple integral
\eq{Stau-m-rep} for the emptiness formation probability takes the
form
\be{Stausym}
\tau(m)=\frac1{m!}\int_C
\frac{Z_m(\{\lambda\},\{\xi\})}
{\prod\limits_{a<b}^m
\sinh(\xi_a-\xi_b)}
{\det}_m[\rho(\lambda_j,\xi_k)]\,d^m\lambda.
\ee
On the other hand, taking the limit $\beta\to\infty$ in \eq{GFebQm+1}, we
obtain
\ba{SebQm+1}
&&{\dis\hspace{1mm}
\lim_{\beta\to\infty}e^{-\beta m}
\langle\exp(\beta Q_{1,m})\rangle=
\frac1{(m!)^2}\oint\limits_\Gamma \prod_{j=1}^n\frac{dz_j}{2\pi i}
\int_C d^n\lambda
\prod_{b=1}^n\prod_{a=1}^{m} \frac{f(z_b,\xi_a)}
{ f(\lambda_b,\xi_a)} }\non
&&{\dis\hspace{21mm}
\times W_m(\{\lambda\},\{z\})\cdot
{\det}_m[t(\lambda_j,z_k)]
\cdot{\det}_m\Bigl[\rho(\lambda_j,z_k)\Bigr].}
\ea
Taking the contour integrals with respect to all $z_j$, we immediately
arrive at \eq{Stausym}.

%
\end{document}